\def \inparg{\leftskip = 40pt\rightskip = 40pt}
\def \outparg{\leftskip = 0 pt\rightskip = 0pt}

\def\npb{{Nucl.\ Phys.\ }{\bf B}}
\def\plb{{Phys.\ Lett.\ }{ \bf B}}

\def\prd{{Phys.\ Rev.\ }{\bf D}}

\def\prl{Phys.\ Rev.\ Lett.\ }

\def\Ttil{\tilde T}
\def\Ahat{\hat A}
\def\Chat{\hat C}

\def\frak#1#2{{\textstyle{{#1}\over{#2}}}}

\def\go{\rightarrow}
\def\lambdabar{\bar\lambda}
\def\lambdahatbar{\bar{\hat\lambda}}
\def\lambdahat{\hat\lambda}

\def\Dhat{\hat D}
\def\Ftil{\tilde F}
\def\Fhat{\hat F}
\def\Atil{\tilde A}
\def\sigmabar{\bar\sigma}
\def\phibar{\bar\phi}
\def\psibar{\bar\psi}
\def\Fbar{\bar F}

\def\Ttil{\tilde T}

\def\dtil{\tilde d}

\def\phitil{\tilde\phi}
\def\psitil{\tilde\psi}
\def\Ncal{{\cal N}}
\def\Ftil{\tilde F}
\def\alphadot{\dot\alpha}
\def\betadot{\dot\beta}

\def\pa{\partial}

\input harvmac
\input epsf
% +--------------------------------------------------------------------+
% |                                                                    |
% |                           TABLES.TEX                               |
% |                                                                    |
% |                     Ray F. Cowan  15-Feb-85                        |
% |                                                                    |
% |                       Princeton University                         |
% |                                                                    |
% |          Present Address:  Laboratory for Nuclear Science          |
% |                            M.I.T.                                  |
% |                            Cambridge, MA 02139                     |
% |                                                                    |
% |                   E-mail:  rfc@slacvm.slac.stanford.edu            |
% |                                                                    |
% |                                                                    |
% |                     Last Revision: 17-Apr-86                       |
% |                                                                    |
% |   Macros I find handy for making tables.  See TABLEDOC TEX for     |
% |   a longer description.  The token-counting macros are straight    |
% |   from the TeXbook's "Dirty Tricks" appendix.                      |
% |                                                                    |
% +--------------------------------------------------------------------+
%
\newbox\hdbox%
\newcount\hdrows%
\newcount\multispancount%
\newcount\ncase%
\newcount\ncols% This is the number of primary text columns in the table.
\newcount\nrows%
\newcount\nspan%
\newcount\ntemp%
\newdimen\hdsize%
\newdimen\newhdsize%
\newdimen\parasize%
\newdimen\spreadwidth%
\newdimen\thicksize%
\newdimen\thinsize%
\newdimen\tablewidth%
\newif\ifcentertables%
\newif\ifendsize%
\newif\iffirstrow%
\newif\iftableinfo%
\newtoks\dbt%
\newtoks\hdtks%
\newtoks\savetks%
\newtoks\tableLETtokens%
\newtoks\tabletokens%
\newtoks\widthspec%
%
%  Book-keeping stuff--see how often these macros are called.
%
%  MOD RFC 900221.
%  Removed usage logging:  it's too complicated under VM/XA.
%\immediate\write15{%
%CP SMSG GJMSINK TEXTABLE --> TABLE MACROS V. 851121 JOB = \jobname%
%}%
%
%  Turn on table diagnostics.
%
\tableinfotrue%
\catcode`\@=11%  Allows use of "@" in macro names, like PLAIN.TEX does.
%  Debugging aid.  Writes #1 on the
%                                    user's terminal and in the log file.
%
%  Define the \tstrut height, depth in terms of the x_height parameter.
%
\def\tstrut{\vrule height3.1ex depth1.2ex width0pt}%
\def\and{\char`\&}%  Allows us to get an `&' in the text.  This is the
%                    same as using the PLAIN TeX macro \&.
\def\tablerule{\noalign{\hrule height\thinsize depth0pt}}%
\thicksize=1.5pt%  Default thickness for fat rules.  The user should feel
%                  free to change this to his preference.
\thinsize=0.6pt%   Default thickness for thin rules.
\def\thickrule{\noalign{\hrule height\thicksize depth0pt}}%
\def\ctr#1{\hfil\ #1\hfil}%
%
%
%
%  Here are things for controlling the width of the finished table.
%
\tablewidth=-\maxdimen%
\spreadwidth=-\maxdimen%
\def\tabskipglue{0pt plus 1fil minus 1fil}%
%
%  Stuff for centering or not.
%
\centertablestrue%
%
%
%
%  \vctr vertically centers its argument in the row.
%
\parasize=4in%
\gdef\ARGS{########}%  Produces the correct number of #'s in the preamble
%                      by the time eveything is expanded and \halign sees
%                      it.
\gdef\headerARGS{####}%  Same as \ARGS, but used in \header macros.
\def\@mpersand{&}%  Allows us to get alignment tab characters later
%                   when we have made the character "&" an active macro.
{\catcode`\|=13%  Make |'s locally active.
\gdef\letbarzero{\let|0}%  Globally define a macro that allows us to
%                          keep active |'s from being expanded in edef's.
\gdef\letbartab{\def|{&&}}%
\gdef\letvbbar{\let\vb|}%
%  This \def will cause active |'s read by
%                            \ruledtable to be converted into double
%                            alignment tabs.
}%  End of locally active |'s.
{\catcode`\&=4%  Make these alignment tabs.
\def\ampskip{&\omit\hfil&}%  This local macro skips a vertical rule.
\catcode`\&=13%  Now make &'s into active macros.
\let&0%  This allows us to expand \ampskip in the next \xdef without
%        attempting to expand the & and getting an "undefined control
%        sequence" error.
\xdef\letampskip{\def&{\ampskip}}%
\gdef\letnovbamp{\let\novb&\let\tab&}
%  This will cause active &'s read by
%                                   \ruledtable to be converted into
%                                   double tabs and an \omit'ted \vrule.
}%  End of locally active &'s.
\def\begintable{%  Here we make |'s and &'s active characters so we can
%                  interpret them as macros.  Note that this action is
%                  true only until we encounter the matching \endgroup
%                  token later at the end of the \ruledtable macro.
   \begingroup%
   \catcode`\|=13\letbartab\letvbbar%
   \catcode`\&=13\letampskip\letnovbamp%
   \def\multispan##1{%  We must redefine \multispan to count the number
%                       of primary columns, not physical columns.
      \omit \mscount##1%
      \multiply\mscount\tw@\advance\mscount\m@ne%
      \loop\ifnum\mscount>\@ne \sp@n\repeat%
   }%  End of \multispan macro.
   \def\|{%
      &\omit\widevline&%
   }%
   \ruledtable%  Now we call \ruledtable to do the real work.
}%  End of \begintable macro.
\long\def\ruledtable#1\endtable{%
%
%  This macro reads in the user's data entries
%  and converts them into a ruled table.
%
%  Important note:  Many macros and parameters are re-defined here, and
%  these must be kept local to the table macros to avoid conflict with
%  their use outside of tables.  This is done by the \begingroup token
%  macro \begintable and the \endgroup token at the end of
%  this macro.
%
   \offinterlineskip%  Needed to make rules touch each other.
   \tabskip 0pt%  Needed for same reason as \offinterlineskip.
   \def\widevline{\vrule width\thicksize}%  Make outer \vrule's wider.
   \def\endrow{\@mpersand\omit\hfil\crnorm\@mpersand}%
   \def\crthick{\@mpersand\crnorm\thickrule\@mpersand}%
   \def\crthickneg##1{\@mpersand\crnorm\thickrule
          \noalign{{\skip0=##1\vskip-\skip0}}\@mpersand}%
   \def\crnorule{\@mpersand\crnorm\@mpersand}%
   \def\crnoruleneg##1{\@mpersand\crnorm
          \noalign{{\skip0=##1\vskip-\skip0}}\@mpersand}%
   \let\nr=\crnorule%  A shorter abbreviation.
   \def\endtable{\@mpersand\crnorm\thickrule}%
   \let\crnorm=\cr%  Allows us to use \cr for our own purposes.
%
%  Cause user-typed \cr's to follow a row with a \tablerule.
%
   \edef\cr{\@mpersand\crnorm\tablerule\@mpersand}%
   \def\crneg##1{\@mpersand\crnorm\tablerule
          \noalign{{\skip0=##1\vskip-\skip0}}\@mpersand}%
   \let\ctneg=\crthickneg
   \let\nrneg=\crnoruleneg
   \the\tableLETtokens%  Get the user's extra \let's, if any.
%
%  Put the data entries into a token register so we can scan through them
%  and see what the user is asking us to do.
%
   \tabletokens={&#1}%  We add an extra alignment tab to the beginning
%                       of the first row to allow for the first \vrule.
%
%  Now count how many rows are in the table and return the result in
%  count register \nrows; do the same for columns, and return that
%  in register \ncols.
%
   \countROWS\tabletokens\into\nrows%
   \countCOLS\tabletokens\into\ncols%
%
%  Now do a little arithmetic to convert the number of primary columns
%  into the number of physical columns that the alignment preamble must
%  prepare for;  similarly for rows.
%
   \advance\ncols by -1%
   \divide\ncols by 2%
   \advance\nrows by 1%
%
%  Tell the user how many rows and columns we found in his data, if he
%  wants to know.
%
   \iftableinfo %
      \immediate\write16{[Nrows=\the\nrows, Ncols=\the\ncols]}%
   \fi%
%
%  Now we actually go ahead and produce the table.
%
   \ifcentertables
      \ifhmode \par\fi%  Make sure we are in vertical mode.
      \line{%  The final table comes out as an \hbox of width the \hsize.
      \hss%  The final table will be centered left-to-right.
   \else %
      \hbox{%
   \fi
      \vbox{%
         \makePREAMBLE{\the\ncols}%  Generate the preamble.
         \edef\next{\preamble}%  This line and the next line force the
         \let\preamble=\next%    expansion of all \ARGS tokens into the
%                                appropriate number of #'s.
         \makeTABLE{\preamble}{\tabletokens}%  Go do the \halign here.
      }%  End of \vbox.
      \ifcentertables \hss}\else }\fi%  Finish the centering effect.
%                                       It is important that no spaces
%                                       follow the two `}' here.
%  }%  End of \line.
   \endgroup%  Return all local macros and parameters to their outside
%              values.
   \tablewidth=-\maxdimen%  Reset \tablewidth to normal.
   \spreadwidth=-\maxdimen% Same for \spreadwidth.
}%  End of macro \ruledtable.
\def\makeTABLE#1#2{%  Does an \halign for the \ruledtable macro.
   {%  Start of local parameter values.
   \let\ifmath0%     These macros would cause trouble if they were to be
   \let\header0%     expanded in the following \xdef; we \let them be
   \let\multispan0%  equal to a digit, because digits can't be expanded.
%
%  Set up the width specification here.
%
   \ncase=0%
   \ifdim\tablewidth>-\maxdimen \ncase=1\fi%
   \ifdim\spreadwidth>-\maxdimen \ncase=2\fi%
   \relax%  This \relax is absolutely necessary, without it the following
%           \ifcase will always take \ncase=0.
%
   \ifcase\ncase %
      \widthspec={}%
   \or %
      \widthspec=\expandafter{\expandafter t\expandafter o%
                 \the\tablewidth}%
   \else %
      \widthspec=\expandafter{\expandafter s\expandafter p\expandafter r%
                 \expandafter e\expandafter a\expandafter d%
                 \the\spreadwidth}%
   \fi %
%\out{Widthspec=[\the\widthspec]}%
%\out{Preamble=[\preamble]}%
   \xdef\next{%  We must force the preamble to be expanded BEFORE the
      \halign\the\widthspec{%
%        \halign is done;  this \edef\next{...}\next construction
%                does the trick.
      #1%  This is the preamble text.
      \noalign{\hrule height\thicksize depth0pt}%  Makes the top \hrule.
      \the#2\endtable%  This is the main body.
%
%     \noalign{\hrule height0.7pt depth0pt}%  Makes the last \hrule.
      }%  End of \halign.
   }%  End of \next.
   }%  End of local values.
   \next%  This \next must be outside of the local values, because now
%          we want those troublesome macros in the \let's above to have
%          their normal actions.
}%  End of macro \makeTABLE.
\def\makePREAMBLE#1{%  This macro generates the necessary preamble for a
%                      ruled table with #1 primary columns.
%                      (Primary columns means the number of columns NOT
%                       counting those used for vertical rules.)
   \ncols=#1%  Get the number of columns desired.
   \begingroup%  Start local parameter definitions.
   \let\ARGS=0%  This is the key to the whole thing; it prevents \ARGS
%                from being expanded in the following \edef's.
   \edef\xtp{\widevline\ARGS\tabskip\tabskipglue%
   &\ctr{\ARGS}\tstrut}%  A 1-column preamble.  Gets the sizing right.
   \advance\ncols by -1%  One column has been generated; decrement the
%                         counter.
   \loop%  Append as many further columns as needed to the preamble.
      \ifnum\ncols>0 %
      \advance\ncols by -1%
      \edef\xtp{\xtp&\vrule width\thinsize\ARGS&\ctr{\ARGS}}%
   \repeat
   \xdef\preamble{\xtp&\widevline\ARGS\tabskip0pt%
   \crnorm}%  Adds the last \vrule.
   \endgroup%  End of local parameters.
}%  End of macro \makePREAMBLE.
\def\countROWS#1\into#2{%  This counts the number of rows in #1 by
%                          looking for control sequences that end a row,
%                          e.g., \cr, \crthick, etc., and puts the result
%                          into count register #2.
   \let\countREGISTER=#2%
   \countREGISTER=0%
%  \out{In countROWS:  tokens are [\the#1]}%
   \expandafter\ROWcount\the#1\endcount%
}%
\def\ROWcount{%
   \afterassignment\subROWcount\let\next= %
}%
\def\subROWcount{%
%  \out{In subROWcount:  next is [\meaning\next]}%  Debugging aid.
   \ifx\next\endcount %
      \let\next=\relax%
   \else%
      \ncase=0%
      \ifx\next\cr %
         \global\advance\countREGISTER by 1%
         \ncase=0%
      \fi%
      \ifx\next\endrow %
         \global\advance\countREGISTER by 1%
         \ncase=0%
      \fi%
      \ifx\next\crthick %
         \global\advance\countREGISTER by 1%
         \ncase=0%
      \fi%
      \ifx\next\crnorule %
         \global\advance\countREGISTER by 1%
         \ncase=0%
      \fi%
      \ifx\next\crthickneg %
         \global\advance\countREGISTER by 1%
         \ncase=0%
      \fi%
      \ifx\next\crnoruleneg %
         \global\advance\countREGISTER by 1%
         \ncase=0%
      \fi%
      \ifx\next\crneg %
         \global\advance\countREGISTER by 1%
         \ncase=0%
      \fi%
      \ifx\next\header %
%     \out{In subROWcount:  next=header, ncase set=1}%
         \ncase=1%
      \fi%
%     \out{In subROWcount:  ncase is [\the\ncase]}%
      \relax%
      \ifcase\ncase %
         \let\next\ROWcount%
%        \out{subROWcount---> ncase=\the\ncase}%
      \or %
         \let\next\argROWskip%
%        \out{subROWcount---> ncase=\the\ncase}%
      \else %
      \fi%
   \fi%
%  \out{subROWcount---> NEXT=\meaning\next}%
   \next%
}%  End of macro \subROWcount.
\def\counthdROWS#1\into#2{%
\dvr{10}%
   \let\countREGISTER=#2%
   \countREGISTER=0%
\dvr{11}%
%  \out{In counthdROWS:  tokens are [\the#1]}%
\dvr{13}%
   \expandafter\hdROWcount\the#1\endcount%
\dvr{12}%
}%
\def\hdROWcount{%
   \afterassignment\subhdROWcount\let\next= %
}%
\def\subhdROWcount{%
%\out{In subhdROWcount:  next is [\meaning\next]}%
   \ifx\next\endcount %
      \let\next=\relax%
   \else%
      \ncase=0%
      \ifx\next\cr %
         \global\advance\countREGISTER by 1%
         \ncase=0%
      \fi%
      \ifx\next\endrow %
         \global\advance\countREGISTER by 1%
         \ncase=0%
      \fi%
      \ifx\next\crthick %
         \global\advance\countREGISTER by 1%
         \ncase=0%
      \fi%
      \ifx\next\crnorule %
         \global\advance\countREGISTER by 1%
         \ncase=0%
      \fi%
      \ifx\next\header %
%\out{In subhdROWcount:  next=header, ncase set=1}%
         \ncase=1%
      \fi%
%\out{In subhdROWcount:  ncase is [\the\ncase]}%
\relax%
      \ifcase\ncase %
         \let\next\hdROWcount%
%\out{subhdROWcount---> ncase=\the\ncase}%
      \or%
         \let\next\arghdROWskip%
%\out{subhdROWcount---> ncase=\the\ncase}%
      \else %
      \fi%
   \fi%
%\out{subhdROWcount---> NEXT=\meaning\next}%
   \next%
}%
{\catcode`\|=13\letbartab
\gdef\countCOLS#1\into#2{%
%  \out{In countCOLS:  tokens are [\the#1]}
   \let\countREGISTER=#2%
   \global\countREGISTER=0%
   \global\multispancount=0%
   \global\firstrowtrue
   \expandafter\COLcount\the#1\endcount%
   \global\advance\countREGISTER by 3%
   \global\advance\countREGISTER by -\multispancount
%  \out{countCOLS-->[\the\countREGISTER]}
}%
\gdef\COLcount{%
   \afterassignment\subCOLcount\let\next= %
}%
{\catcode`\&=13%
\gdef\subCOLcount{%
%\out{In subCOLcount: next is [\meaning\next]}
   \ifx\next\endcount %
      \let\next=\relax%
   \else%
      \ncase=0%
      \iffirstrow
         \ifx\next& %
            \global\advance\countREGISTER by 2%
            \ncase=0%
         \fi%
         \ifx\next\span %
            \global\advance\countREGISTER by 1%
            \ncase=0%
         \fi%
         \ifx\next| %
            \global\advance\countREGISTER by 2%
            \ncase=0%
         \fi
         \ifx\next\|
            \global\advance\countREGISTER by 2%
            \ncase=0%
         \fi
         \ifx\next\multispan
            \ncase=1%
            \global\advance\multispancount by 1%
         \fi
         \ifx\next\header
            \ncase=2%
         \fi
         \ifx\next\cr       \global\firstrowfalse \fi
         \ifx\next\endrow   \global\firstrowfalse \fi
         \ifx\next\crthick  \global\firstrowfalse \fi
         \ifx\next\crnorule \global\firstrowfalse \fi
         \ifx\next\crnoruleneg \global\firstrowfalse \fi
         \ifx\next\crthickneg  \global\firstrowfalse \fi
         \ifx\next\crneg       \global\firstrowfalse \fi
      \fi%  End of \iffirstrow.
\relax%\out{subCOL-->  ncase=[\the\ncase]}
% \out{subCOL-->  next=\meaning\next}
      \ifcase\ncase %
         \let\next\COLcount%
      \or %
         \let\next\spancount%
      \or %
         \let\next\argCOLskip%
      \else %
      \fi %
   \fi%
%  \out{subCOL-->  countREGISTER=[\the\countREGISTER]}
   \next%
}%
\gdef\argROWskip#1{%
%  Deletes the next balanced, undelimited argument from a
%                 token list.
% \out{---> Entering argROWskip <---}
% \out{In argROWskip:  deleted arg is [#1]}%
   \let\next\ROWcount \next%
}%  End of macro \argskip.
\gdef\arghdROWskip#1{%
%  Deletes the next balanced, undelimited argument from a
%                 token list.
% \out{---> Entering arghdROWskip <---}
% \out{In arghdROWskip:  deleted arg is [#1]}%
   \let\next\ROWcount \next%
}%  End of macro \arghdROWskip.
\gdef\argCOLskip#1{%
%  Deletes the next balanced, undelimited argument from a
%                 token list.
% \out{---> Entering argCOLskip <---}
% \out{In argCOLskip:  deleted arg is [#1]}%
   \let\next\COLcount \next%
}%  End of macro \argskip.
}%  End of active &'s.
}%  End of active |'s.
\def\spancount#1{%\out{spancount--->\meaning#1}
   \nspan=#1\multiply\nspan by 2\advance\nspan by -1%
   \global\advance \countREGISTER by \nspan
%  \out{number spancount--->\the\nspan; \the\countREGISTER}
   \let\next\COLcount \next}%
\def\dvr#1{\relax}%
% \omit\hfil%
% \parindent=0pt\hsize=1.1in\valign{%
% \vfil#\vfil&\vfil#\vfil\cr\hfil\hbox{\ Added to\ }\hfil&%
% \hfil\hbox{\ empty events\ }\hfil\cr}\hfil%
\def\header#1{%
\dvr{1}{\let\cr=\@mpersand%
\hdtks={#1}%
%\out{In header:  hdtks=[\the\hdtks]}%
\counthdROWS\hdtks\into\hdrows%
\advance\hdrows by 1%
\ifnum\hdrows=0 \hdrows=1 \fi%
%\out{In header:  Nhdrows=[\the\hdrows]}%
\dvr{5}\makehdPREAMBLE{\the\hdrows}%
%\out{In header:  headerpreamble=[\headerpreamble]}%
\dvr{6}\getHDdimen{#1}%
%\out{In header:  hdsize=[\the\hdsize]}%
%\striplastCR{#1}%
{\parindent=0pt\hsize=\hdsize{\let\ifmath0%
\xdef\next{\valign{\headerpreamble #1\crnorm}}}\dvr{7}\next\dvr{8}%
}%
}\dvr{2}}%  End of macro \header.
\def\makehdPREAMBLE#1{%This macro generates the necessary preamble for a
\dvr{3}%
%                      ruled table with \ncols primary columns.
%                      (Primary columns means the number of columns NOT
%                       counting those used for vertical rules.
\hdrows=#1%  Get the number of columns desired.
{%  Start local parameter definitions.
\let\headerARGS=0%
%  This is the key to the whole thing; it prevents \ARGS
\let\cr=\crnorm%
%                from being expanded in the followin \edef's.
\edef\xtp{\vfil\hfil\hbox{\headerARGS}\hfil\vfil}%
\advance\hdrows by -1%  One row has been generated; decrement the
%                         counter.
\loop%  Append as many further rows as needed to the preamble.
\ifnum\hdrows>0%
\advance\hdrows by -1%
\edef\xtp{\xtp&\vfil\hfil\hbox{\headerARGS}\hfil\vfil}%
\repeat%
\xdef\headerpreamble{\xtp\crcr}%
}%  End of local parameters.
\dvr{4}}%  End of \makehdPREAMBLE.
\def\getHDdimen#1{%
%\out{In getHDdimen:  Arg 1=[#1]}%
\hdsize=0pt%
\getsize#1\cr\end\cr%
}%  End of macro getHDdimen.
\def\getsize#1\cr{%
%\out{In getsize:  Arg 1=[#1]}%
%  Here we have to check arg#1 and see if the first token in #1 is an
%    \end; if so, we stop, else we check the width of arg#1.
%  We recall that each arg#1 will be terminated with a \cr token.
\endsizefalse\savetks={#1}%
%\out{In getsize:  the savetks = [\the\savetks]}%
\expandafter\lookend\the\savetks\cr%
%\out{In getsize:  ifendsize = [\meaning\ifendsize]}%
\relax \ifendsize \let\next\relax \else%
\setbox\hdbox=\hbox{#1}\newhdsize=1.0\wd\hdbox%
\ifdim\newhdsize>\hdsize \hdsize=\newhdsize \fi%
%\out{In getsize:  hdsize=[\the\hdsize]}%
%\out{In getsize:  newhdsize=[\the\newhdsize]}%
\let\next\getsize \fi%
\next%
}%
\def\lookend{\afterassignment\sublookend\let\looknext= }%
\def\sublookend{\relax%
%\out{In sublookend:  looknext = [\looknext]}%
\ifx\looknext\cr %
%\out{In sublooknext:  looknext=cr}%
\let\looknext\relax \else %
%\out{In sublooknext:  looknext/=cr}%
   \relax
   \ifx\looknext\end \global\endsizetrue \fi%
   \let\looknext=\lookend%
    \fi \looknext%
}%
%
%  Allow the user to make his own names for crthick, etc.
%
\def\tablelet#1{%
   \tableLETtokens=\expandafter{\the\tableLETtokens #1}%
}%
\catcode`\@=12%  Change @'s back to their normal category code.

{\nopagenumbers
\line{\hfil LTH 653}
\line{\hfil hep-th/0505248}
%\line{\hfil Revised Version}
\vskip .5in
\centerline{\titlefont One-loop renormalisation of general}
\centerline{\titlefont $\Ncal=\frak12$ supersymmetric gauge theory}
%\medskip
\vskip 1in
\centerline{\bf I.~Jack, D.R.T.~Jones and L.A.~Worthy}
\bigskip
\centerline{\it Department of Mathematical Sciences,  
University of Liverpool, Liverpool L69 3BX, U.K.}
\vskip .3in
We investigate the one-loop renormalisability of a general $\Ncal=\frak12$ 
supersymmetric gauge theory coupled to chiral matter, and show the 
existence of an $\Ncal=\frak12$ supersymmetric $SU(N)\times U(1)$ theory which 
is renormalisable at one loop.  
\Date{May 2005}}

\newsec{Introduction} There has recently been much interest in theories
defined on non-anti-commutative superspace \ref\seib{N.~Seiberg, JHEP
{\bf 0306} (2003) 010} \ref\araki{T.~Araki, K.~Ito and  A. Ohtsuka,
\plb573 (2003) 209}. Such theories are non-hermitian and turn out to
have only half the supersymmetry of the corresponding ordinary
supersymmetric theory--hence the term ``$\Ncal=\frak12$ supersymmetry''. 
These theories are not power-counting  renormalisable but it has been
argued\ref\terash{S. Terashima and J-T Yee, JHEP {\bf 0312} 
(2003) 053}\nref\gris{M.T.~Grisaru, S.~Penati and  A.~Romagnoni, JHEP {\bf
0308} (2003) 003\semi
R.~Britto and B.~Feng, \prl91 (2003) 201601\semi
A.~Romagnoni, JHEP {\bf 0310} (2003) 016}\nref\lunin{O.~Lunin 
and S.-J. Rey, JHEP  {\bf 0309}
(2003) 045}\nref\alish{
M.~Alishahiha, A.~Ghodsi and N.~Sadooghi, \npb691
(2004) 111}--\ref\berrey{D.~Berenstein and S.-J.~Rey, \prd68 (2003) 121701}
 that they are in  fact nevertheless
renormalisable, in the sense that only a finite number of additional
terms need to be added to the lagrangian to absorb divergences to all
orders. This is primarily because although the theory contains operators
of dimension five and higher, they are not accompanied by their
hermitian conjugates which would be required to generate divergent
diagrams. This argument does not of course guarantee that the precise 
form of the 
lagrangian will be preserved by renormalisation; nor does the $\Ncal=\frak12$ 
supersymmetry, since some terms in the lagrangian are inert under this
symmetry.    
Moreover, the argument also requires (in the gauged case) the
assumption of gauge invariance to rule out some classes of divergent 
structure. As we showed in Ref.~\ref\jjw{I.~Jack, D.R.T.~Jones 
and L.A.~Worthy, \plb611 (2005) 199}, there are
problems with this assumption; even at one loop, at least in the
standard class of gauges, divergent non-gauge-invariant terms are
generated. However, in the case of pure $\Ncal=\frak12$
supersymmetry (i.e. no chiral matter) we displayed a divergent field 
redefinition
which miraculously removed the non-gauge-invariant terms and restored
gauge invariance. Moreover, we displayed a slightly modified
(but still $\Ncal=\frak12$ supersymmetric) 
version of the original pure $\Ncal=\frak12$ 
lagrangian which had a form preserved 
under renormalisation. The authors of Ref.~\ref\penrom{
S.~Penati and A.~Romagnoni, JHEP {\bf 0502} (2005) 064}
obtained the one loop effective action for pure $\Ncal=\frak12$
supersymmetry using a superfield formalism. Although they found
divergent contributions which broke supergauge invariance, their final
result was gauge-invariant without the need for any redefinition. On the
other hand it is hard to make any inferences about renormalisability from 
their superfield form of the one-loop result.
In the present work we consider the $\Ncal=\frak12$
supersymmetric action coupled to chiral matter. 
The original non-anticommutative theory defined in 
superfields appears to require a $U(N)$ gauge group\araki\terash. In 
Ref.~\jjw\ we 
considered  the component form of the pure $\Ncal=\frak12$
supersymmetric action adapted to $SU(N)$.
We argued that it was only for $SU(N)$ that a form-invariant lagrangian
could be defined; indeed the $U(N)$ gauge symmetry is not preserved under
renormalisation. In the case with chiral matter it turns out that 
the lagrangian is no longer form-invariant in the $SU(N)$ case either. In fact,
a general $\Ncal=\frak12$ supersymmetric $SU(N)$ invariant action cannot be defined.
However, we shall demonstrate the existence of a new $\Ncal=\frak12$
supersymmetric $SU(N)\times U(1)$ action which is renormalisable and 
preserves $\Ncal=\frak12$ supersymmetry at one loop. 

The action for an $\Ncal=\frak12$ supersymmetric $U(N)$ gauge theory coupled to 
chiral matter is given by\araki
\eqn\lagrana{\eqalign{
 S=&\int d^4x
\Bigl[\tr\{-\frak12F^{\mu\nu}F_{\mu\nu}-2i\lambdabar\sigmabar^{\mu}
(D_{\mu}\lambda)+D^2\}\cr
&-2igC^{\mu\nu}\tr\{F_{\mu\nu}\lambdabar\lambdabar\}
+g^2|C|^2\tr\{(\lambdabar\lambdabar)^2\}\cr
&+\Bigl\{
\Fbar F -i\psibar\sigmabar^{\mu}D_{\mu}\psi-D^{\mu}\phibar D_{\mu}\phi\cr
&+g\phibar D\phi +i\sqrt2g(\phibar \lambda\psi-\psibar\lambdabar\phi)\cr
&+\sqrt2gC^{\mu\nu}D_{\mu}\phibar\lambdabar\sigmabar_{\nu}\psi
+igC^{\mu\nu}\phibar F_{\mu\nu}F
+\frak{1}{4}|C|^2g^2\phibar\lambdabar\lambdabar
F\cr
&+(\phi\go\phitil,\psi\go\psitil,F\go\Ftil,C^{\mu\nu}\go
-C^{\mu\nu})\Bigr\}\Bigr],\cr}}
where we include a multiplet $\{\phi,\psi,F\}$ transforming according to the
fundamental representation and, to ensure anomaly cancellation, a 
multiplet $\{\phitil,\psitil,\Ftil\}$ transforming according to its conjugate.
We define
\eqn\afdef{
D_{\mu}\phi=\pa_{\mu}\phi+igA_{\mu}\phi,
\quad D_{\mu}\lambda=\pa_{\mu}\lambda+ig[A_{\mu},\lambda],
\quad F_{\mu\nu}=\pa_{\mu}A_{\nu}-\pa_{\nu}A_{\mu}+ig[A_{\mu},A_{\nu}],}
(with a similar expression for $D_{\mu}\phitil$)
where
\eqn\gpex{
A_{\mu}=A_{\mu}^AR^A,\quad \lambda=\lambda^AR^A,\quad D=D^AR^A,}
with $R^A$ being the group matrices for $U(N)$ in the fundamental 
representation. These satisfy
\eqn\comrel{ [R^A,R^B]=if^{ABC}R^C,\quad
\{R^A,R^B\}=d^{ABC}R^C,}
where $d^{ABC}$ is totally symmetric. If one decomposes $U(N)$ as 
$SU(N)\times U(1)$ then our convention is that $R^a$ are the $SU(N)$
generators and $R^0$ the $U(1)$ generator. Of course then $f^{ABC}=0$ 
unless all indices are $SU(N)$. The matrices are normalised so that
$\Tr[R^AR^B]=\frak12\delta^{AB}$. In particular, $R^0=\sqrt{\frak{1}{2N}}1$.
We note that $d^{ab0}=\sqrt{\frak2N}\delta^{ab}$, $d^{000}=\sqrt{\frak2N}$. 
In Eq.~\lagrana, $C^{\mu\nu}$ is related to the non-anti-commutativity 
parameter $C^{\alpha\beta}$ by  
\eqn\Cmunu{
C^{\mu\nu}=C^{\alpha\beta}\epsilon_{\beta\gamma}
\sigma^{\mu\nu}_{\alpha}{}^{\gamma},} 
where 
\eqn\sigmunu{\eqalign{
\sigma^{\mu\nu}=&\frak14(\sigma^{\mu}\sigmabar^{\nu}-
\sigma^{\nu}\sigmabar^{\mu}),\cr
\sigmabar^{\mu\nu}=&\frak14(\sigmabar^{\mu}\sigma^{\nu}-
\sigmabar^{\nu}\sigma^{\mu}),\cr }} 
and 
\eqn\Csquar{
|C|^2=C^{\mu\nu}C_{\mu\nu}.} 
Our conventions are in accord with \seib; in particular, 
\eqn\sigid{
\sigma^{\mu}\sigmabar^{\nu}=-\eta^{\mu\nu}+2\sigma^{\mu\nu}.}
Properties of $C$ which follow from
Eq.~\Cmunu\ are  
\eqna\cprop$$\eqalignno{
C^{\alpha\beta}&=\frak12\epsilon^{\alpha\gamma}
\left(\sigma^{\mu\nu}\right)_\gamma{}^{\beta}C_{\mu\nu},
& \cprop a\cr
C^{\mu\nu}\sigma_{\nu\alpha\betadot}&=C_{\alpha}{}^{\gamma}
\sigma^{\mu}{}_{\gamma\betadot},&\cprop b\cr
C^{\mu\nu}\sigmabar_{\nu}^{\alphadot\beta}&=-C^{\beta}{}_{\gamma}
\sigmabar^{\mu\alphadot\gamma}.&\cprop c\cr}$$ 

Upon substituting Eq.~\gpex\
into Eq.~\lagrana\ and using Eq.~\comrel, we obtain
the action in the $U(N)$ case in the form:
\eqn\lagran{\eqalign{
 S=&\int d^4x
\Bigl[-\frak14F^{\mu\nu A}F^A_{\mu\nu}-i\lambdabar^A\sigmabar^{\mu}
(D_{\mu}\lambda)^A+\frak12D^AD^A\cr
&-\frak12igC^{\mu\nu}d^{ABC}F^A_{\mu\nu}\lambdabar^B\lambdabar^C
+\frak18g^2|C|^2d^{ABE}d^{CDE}(\lambdabar^A\lambdabar^B)
(\lambdabar^C\lambdabar^D)\cr
&+\Bigl\{
\Fbar F -i\psibar\sigmabar^{\mu}D_{\mu}\psi-D^{\mu}\phibar D_{\mu}\phi\cr
&+g\phibar D\phi +i\sqrt2g(\phibar \lambda\psi-\psibar\lambdabar\phi)\cr
&+\sqrt2gC^{\mu\nu}D_{\mu}\phibar\lambdabar\sigmabar_{\nu}\psi
+igC^{\mu\nu}\phibar F_{\mu\nu}F
+\frak{1}{8}|C|^2g^2d^{ABC}\phibar R^A\lambdabar^B\lambdabar^C
F\cr
&+(\phi\go\phitil,\psi\go\psitil,F\go\Ftil,C^{\mu\nu}\go
-C^{\mu\nu})\Bigr\}\Bigr].\cr}} 
with gauge coupling $g$, gauge field $A_{\mu}$, gaugino $\lambda$ and with
\eqn\fmunu{\eqalign{
F_{\mu\nu}^A=&\pa_{\mu}A_{\nu}^A-\pa_{\nu}A_{\mu}^A-gf^{ABC}A_{\mu}^BA_{\nu}^C
,\cr
D_{\mu}\lambda^A=&\pa_{\mu}\lambda^A-gf^{ABC}A_{\mu}^B\lambda^C.\cr}}
However, it is clear that the 
$U(N)$ action cannot be renormalisable, since for any $U(N)$ gauge theory the
gauge couplings for the $SU(N)$ and $U(1)$ parts of the theory renormalise
differently. To obtain 
a renormalisable theory one must introduce different couplings for the
$SU(N)$ and $U(1)$ parts of the gauge group and then the 
$U(N)$ gauge-invariance is lost. This is a trivial point but one 
which does
not seem to have been made in other discussions of the renormalisation
of $\Ncal=\frak12$ supersymmetric gauge theory. Remarkably, we shall see that by
a judicious introduction of different couplings for the $SU(N)$ and $U(1)$ 
parts of the gauge group, we can obtain an $SU(N)\times U(1)$ theory which
still has $\Ncal=\frak12$ supersymmetry which is preserved under
renormalisation. We propose replacing Eq.~\lagran\ by
\eqn\lagranb{\eqalign{
S=&\int d^4x
\Bigl[-\frak14F^{\mu\nu A}F^A_{\mu\nu}-i\lambdabar^A\sigmabar^{\mu}
(D_{\mu}\lambda)^A+\frak12D^AD^A\cr
&-\frak12iC^{\mu\nu}d^{ABC}e^{ABC}F^A_{\mu\nu}\lambdabar^B\lambdabar^C\cr
&+\frak18g^2|C|^2d^{abe}d^{cde}(\lambdabar^a\lambdabar^b)
(\lambdabar^c\lambdabar^d)
+\frak{1}{4N}\frak{g^4}{g_0^2}|C|^2(\lambdabar^a\lambdabar^a)
(\lambdabar^b\lambdabar^b)\cr
&+\Bigl\{
\Fbar F -i\psibar\sigmabar^{\mu}D_{\mu}\psi-D^{\mu}\phibar D_{\mu}\phi\cr
&+\phibar \Dhat\phi +
i\sqrt2(\phibar \lambdahat\psi-\psibar\lambdahatbar\phi)\cr
&+\sqrt2C^{\mu\nu}D_{\mu}\phibar\lambdahatbar\sigmabar_{\nu}\psi
+iC^{\mu\nu}\phibar \Fhat_{\mu\nu}F
+\frak{1}{8}|C|^2d^{ABC}\phibar R^A\lambdahatbar^B\lambdahatbar^C
F\cr
&+\frak{1}{N}\gamma_1g_0^2|C|^2(\lambdabar^a\lambdabar^a)
(\lambdabar^0\lambdabar^0)\cr
&-\gamma_2 C^{\mu\nu}g\left(
\sqrt2D_{\mu}\phibar\lambdabar^aR^a\sigmabar_{\nu}\psi+\sqrt2\phibar
\lambdabar^aR^a
\sigmabar_{\nu}D_{\mu}\psi+i\phibar F^a_{\mu\nu}R^aF\right)\cr
&+(\phi\go\phitil,\psi\go\psitil,F\go\Ftil,C^{\mu\nu}\go
-C^{\mu\nu})\Bigr\}\Bigr],\cr}} 
where $\gamma_1$, $\gamma_2$ are constants, 
\eqn\hatdefs{
\Ahat_{\mu}=\Ahat_{\mu}^AR^A=gA_{\mu}^aR^a+g_0A_{\mu}^0R^0,}
with similar definitions for $\lambdahat$, $\Dhat$, $\Fhat_{\mu\nu}$,
and now
\eqn\covdefs{
D_{\mu}\phi=(\pa_{\mu}+i\Ahat_{\mu})\phi.}
We also have
\eqn\etensor{
e^{abc}=g,\quad e^{a0b}=e^{ab0}=e^{000}=g_0,\quad e^{0ab}={g^2\over{g_0}}.}
It is easy to show that Eq.~\lagranb\ is invariant under 
\eqn\newsusy{
\eqalign{
\delta A^A_{\mu}=&-i\lambdabar^A\sigmabar_{\mu}\epsilon\cr
\delta \lambda^A_{\alpha}=&i\epsilon_{\alpha}D^A+\left(\sigma^{\mu\nu}\epsilon
\right)_{\alpha}\left[F^A_{\mu\nu}
+\frak12iC_{\mu\nu}e^{ABC}d^{ABC}\lambdabar^B\lambdabar^C\right],\quad
\delta\lambdabar^A_{\alphadot}=0,\cr
\delta D^A=&-\epsilon\sigma^{\mu}D_{\mu}\lambdabar^A,\cr
\delta\phi=&\sqrt2\epsilon\psi,\quad\delta\phibar=0,\cr
\delta\psi^{\alpha}=&\sqrt2\epsilon^{\alpha} F,\quad 
\delta\psibar_{\alphadot}=-i\sqrt2(D_{\mu}\phibar)
(\epsilon\sigma^{\mu})_{\alphadot},\cr
\delta F=&0,\quad
\delta \Fbar=-i\sqrt2D_{\mu}\psibar\sigmabar^{\mu}\epsilon
-2i\phibar\epsilon\lambdahat+2C^{\mu\nu}D_{\mu}(\phibar\epsilon\sigma_{\nu}
\lambdahatbar).\cr}}
Apart from the term with the coefficient $\gamma_1$ and  
the group of terms with coefficient $\gamma_2$, Eq.~\lagranb\ reduces to the
original $U(N)$ lagrangian Eq.~\lagran\ derived from nonanticommuting 
superspace upon setting $g_0=g$. These remaining terms are 
separately invariant under $\Ncal=\frak12$ supersymmetry and must
be included to obtain a renormalisable lagrangian, as we shall see.

In Ref.~\jjw\ we gave an $SU(N)$-invariant theory with $\Ncal=\frak12$ 
supersymmetry in the pure gauge case. The supersymmetry transformations in that
case were essentially obtained by striking out any $0$ index in the $U(N)$
transformations. However in the general case these transformations do not 
close, since the gauge-field part of
$\sqrt2gC^{\mu\nu}D_{\mu}\phibar\lambdabar\sigmabar_{\nu}\psi$ term
produces a $C^{\mu\nu}\phibar\lambdabar^a\lambdabar^a\sigmabar_{\nu}\psi$ term 
which in the $U(N)$ case is cancelled by the variation of 
$\phibar \lambda^0\psi$, a term which is absent for $SU(N)$. 
Of course because of the
${g^2\over{g_0}}$ terms, one cannot obtain the $SU(N)$ theory simply by 
setting $g_0=0$ in the $SU(N)\times U(1)$ theory.  

We use the standard gauge-fixing term 
\eqn\gafix{
S_{\rm{gf}}={1\over{2\alpha}}\int d^4x (\pa.A)^2} 
with its associated
ghost terms.  The gauge propagators for $SU(N)$ and $U(1)$ are both given by  
\eqn\gprop{
\Delta_{\mu\nu}=-{1\over{p^2}}\left(\eta_{\mu\nu}
+(\alpha-1){p_{\mu}p_{\nu}\over{p^2}}\right)}
(omitting group factors) and the gaugino propagator is  
\eqn\fprop{
\Delta_{\alpha\alphadot}={p_\mu\sigma^{\mu}_{\alpha\alphadot}\over{p^2}},}
where the momentum enters at the end of the propagator with the undotted 
index.  
The one-loop graphs contributing
to the ``standard'' terms in the lagrangian (those without a
$C^{\mu\nu}$) are the same as in the ordinary $\Ncal=1$ case, so 
anomalous dimensions and gauge $\beta$-functions are as for
$\Ncal=1$. Since our gauge-fixing term in Eq.~\gafix\ does not preserve 
supersymmetry, the anomalous dimensions for $A_{\mu}$ and $\lambda$
are
different (and moreover gauge-parameter dependent), as are those for
$\phi$ and $\psi$. However, the 
gauge $\beta$-functions are of course gauge-independent. 
The one-loop one-particle-irreducible (1PI) 
graphs contributing to the new terms (those
containing $C$) are depicted in Figs.~1--8. 

\newsec{Renormalisation of the $SU(N)\times U(1)$ action}
Ordinarily the divergences in one-loop diagrams should be cancelled by the 
one-loop divergences in $S_B$, obtained by  
replacing the fields and couplings in Eq.~\lagranb\   
with bare fields and couplings given by    
\eqn\bare{\eqalign{ \lambda^a_B=Z_{\lambda}^{\frak12}\lambda^a,
\quad \lambda^0_B=&Z_{\lambda^0}^{\frak12}\lambda^0,\quad
A^{a}_{\mu B}=Z_A^{\frak12}A^{a}_{\mu},\quad A^{0}_{\mu B}
=Z_{A^0}^{\frak12}A^0_{\mu},\cr  
\phi_B=Z_{\phi}^{\frak12}\phi,
\quad \psi_B=&Z_{\psi}^{\frak12}\psi,\quad g_B=Z_gg,\quad g_{0B}=Z_{g_0}g,\cr
\gamma_{1B}=Z_1, \quad
\gamma_{2B}=&Z_2,\quad C_B^{\mu\nu}=Z_CC^{\mu\nu}, \quad |C|_B^2=Z_{|C|^2}|C|^2.
\cr}} 
In Eq.~\bare, $Z_1$ and $Z_2$ are divergent contributions, in other words
we have set the renormalised couplings $\gamma_1$ and $\gamma_2$ to zero for
simplicity. The other renormalisation constants start with 
tree-level values of 1. As we mentioned before,
the renormalisation constants for the fields
and for the gauge couplings $g$, $g_0$ are the same as in the ordinary $\Ncal=1$
supersymmetric theory\lunin\ and are therefore given up to one loop 
by\ref\timj{
D.~Gross and F.~Wilcek, \prd8 (1973) 3633\semi
D.R.T.~Jones, \npb87 (1975) 127}: 
\eqn\Zgg{\eqalign{
Z_{\lambda}&=1-g^2L(2\alpha N +2),\cr
Z_A&=1+g^2L[(3-\alpha)N-2]\cr
Z_g&=1+g^2L\left(1-3N\right),\cr
Z_{\phi}=&1+2(1-\alpha)L\Chat_2,\cr
Z_{\psi}=&1-2(1+\alpha)L\Chat_2,}}
where (using dimensional regularisation with $d=4-\epsilon$)
$L={1\over{16\pi^2\epsilon}}$ and
\eqn\ctildef{
\Chat_2=g^2R^aR^a+g_0^2R^0R^0=\frak12\left(Ng^2+\frak1N\Delta\right)}
with
\eqn\deltadefn{
\Delta=g_0^2-g^2.}
(We have given here the renormalisation constants corresponding to the
$SU(N)$ sector of the $U(N)$ theory; those for the $U(1)$ sector, namely
$Z_{\lambda^0}$, $Z_{A^0}$ and $Z_{g_0}$, are given by omitting the terms in
$N$ and replacing $g$ by $g_0$.) 

Upon inserting Eq.~\Zgg\ into Eq.~\lagranb\ we obtain the one-loop 
contributions from $S_B$ as 
\eqn\finresun{\eqalign{
S_B^{(1)}=&L\int d^4x\Biggl(iC^{\mu\nu}
\Bigl[\bigl\{\frak12(3+5\alpha)N+2\bigr\}g^3d^{abc}
\pa_{\mu}A^a_{\nu}\lambdabar^b\lambdabar^c\cr
&+[3(\alpha-1)N+4]g^2g_0d^{ab0}\pa_{\mu}A^a_{\nu}\lambdabar^b\lambdabar^0\cr
&+2[(3+\alpha)Ng^2+g_0^2]{g^2\over{g_0}}
d^{0bc}\pa_{\mu}A^0_{\nu}\lambdabar^b\lambdabar^c
+2g_0^2d^{000}\pa_{\mu}A^0_{\nu}\lambdabar^0\lambdabar^0\Bigr]\cr
&-\left[\frak32(1+\alpha)N+1\right]i
d^{abe}f^{cde}g^4C^{\mu\nu}A_{\mu}^cA_{\nu}^d\lambdabar^a\lambdabar^b\cr
&-2i(\alpha N+1)
d^{0be}f^{cde}g^3g_0C^{\mu\nu}A_{\mu}^cA_{\nu}^d\lambdabar^0\lambdabar^b\cr
&-g^2|C|^2\Bigl[\frak14[(3+2\alpha)N+1]g^2
d^{abe}d^{cde}(\lambdabar^a\lambdabar^b)
(\lambdabar^c\lambdabar^d)\cr
&+\left[(3+\alpha)\frak{g^4}{g_0^2}+\frak{g^2}{2N}\right]
(\lambdabar^a\lambdabar^a)(\lambdabar^b\lambdabar^b)
-\frak{Z^{(1)}_1}{N}g_0^2(\lambdabar^a\lambdabar^a)
(\lambdabar^0\lambdabar^0)\Bigr]\cr
&-\frak12iZ_C^{(1)}C^{\mu\nu}d^{ABC}e^{ABC}F^A_{\mu\nu}\lambdabar^B\lambdabar^C\cr
&+Z_{|C|^2}^{(1)}\left[\frak18g^2
|C|^2d^{abe}d^{cde}(\lambdabar^a\lambdabar^b)
(\lambdabar^c\lambdabar^d)
+\frak{1}{4N}\frak{g^4}{g_0^2}|C|^2(\lambdabar^a\lambdabar^a)
(\lambdabar^b\lambdabar^b)\right]\cr
&+\Bigl\{\sqrt2C^{\mu\nu}\Bigl[-\left((3+\alpha)Ng^2+Z^{(1)}_2+2\alpha\Chat_2\right)
g\pa_{\mu}\phibar \lambdabar^aR^a\sigmabar_{\nu}\psi-2\alpha\Chat_2g_0
\pa_{\mu}\phibar \lambdabar^0R^0\sigmabar_{\nu}\psi\cr
&-Z^{(1)}_2g
\phibar \lambdabar^aR^a\sigmabar_{\nu}\pa_{\mu}\psi\Bigr]\cr
&+i\sqrt2C^{\mu\nu}\Bigl[g^2A_{\mu}^b\phibar \lambdabar^b
\Bigl[-Z^{(1)}_2R^aR^b+\left(\frak12Ng^2(9+3\alpha)+Z^{(1)}_2\right)
R^bR^a+2\alpha\Chat_2\Bigl]
\sigmabar_{\nu}\psi\cr
&+gg_0\left(\frak12Ng^2(3+\alpha)+2\alpha\Chat_2\right)A_{\mu}^b\phibar 
\lambdabar^0
R^0R^b\sigmabar_{\nu}\psi\cr
&+gg_0\left((3+\alpha)Ng^2+2\alpha \Chat_2\right)
A_{\mu}^0\phibar \lambdabar^a
R^aR^0\sigmabar_{\nu}\psi
+2g_0^2\alpha \Chat_2A_{\mu}^0\phibar \lambdabar^0
(R^0)^2\sigmabar_{\nu}\psi\Bigr]\cr
&+iC^{\mu\nu}\phibar\Bigl[\left(2(1-\alpha)\Chat_2-(3+\alpha)Ng^2-2Z^{(1)}_2\right)
g\pa_{\mu}A^a_{\nu}R^a
+2(1-\alpha)\Chat_2g_0\pa_{\mu}A^0_{\nu}R^0\cr
&+\left((\alpha-1)\Chat_2+(3+\alpha)Ng^2+Z^{(1)}_2\right)g^2f^{abc}
A^a_{\mu}A^b_{\nu}R^c\Bigr]F\cr
&+\frak{1}{8}|C|^2\Bigl[\left((1-\alpha)\Chat_2-(6+2\alpha)Ng^2\right)
g^2d^{Abc}\phibar R^A\lambdabar^b\lambdabar^c\cr
&+2\left((1-\alpha)\Chat_2-(3+\alpha)Ng^2\right)
g^2d^{a0c}\phibar R^a\lambdabar^0\lambdabar^c
+(1-\alpha)\Chat_2d^{000}\phibar R^0\lambdabar^0\lambdabar^0\Bigr]F\cr
&+Z_C^{(1)}C^{\mu\nu}\left[\sqrt2D_{\mu}\phibar\lambdahatbar\sigmabar_{\nu}\psi
+i\phibar \Fhat_{\mu\nu}F\right]
+\frak{1}{8}Z_{|C|^2}^{(1)}|C|^2d^{ABC}\phibar R^A\lambdahatbar^B\lambdahatbar^C
F\cr
&+\frak{1}{N}\gamma_1g_0^2|C|^2(\lambdabar^a\lambdabar^a)
(\lambdabar^0\lambdabar^0)\cr
&-\gamma_2 C^{\mu\nu}g\left(
\sqrt2D_{\mu}\phibar\lambdabar^aR^a\sigmabar_{\nu}\psi+\sqrt2\phibar
\lambdabar^aR^a
\sigmabar_{\nu}D_{\mu}\psi+i\phibar F^a_{\mu\nu}R^aF\right)\cr
&+(\phi\go\phitil,\psi\go\psitil,F\go\Ftil,C^{\mu\nu}\go
-C^{\mu\nu})\Bigr\}\Biggr).\cr}}
The results $\Gamma_{i\rm{1PI}}^{(1)\rm{pole}}$, $i=1\ldots8$
for the one-loop divergences from the 1PI graphs in Figs. 1--8 respectively
are given in Appendix A. 
It is clear that they cannot be cancelled by Eq.~\finresun,
in particular since they contain contributions involving $\sigmabar^{\mu\nu}$
which do not appear in Eq.~\finresun. 
As we showed in Ref.~\jjw, this can be remedied by field redefinitions, or,
to put it another way, additional non-linear field renormalisations.
We find that a field redefinition 
\eqn\lchange{
\delta\lambda^A=-\frak12NLg^2C^{\mu\nu}e^{BAC}d^{ABC}c^Ac^Bd^C
\sigma_{\mu}\lambdabar^CA_{\nu}^B,}
where $c^A=1-\delta^{A0}$,
$d^A=1+\delta^{A0}$, results in a change in the action
\eqn\delS{\eqalign{
\delta S_{\lambda}=&NLg^2\int d^4x\Bigl[
-\frak14iC^{\mu\nu}(d^{abc}g\pa_{\mu}A_{\nu}^a\lambdabar^b\lambdabar^c
-d^{abe}f^{cde}g^2A_{\mu}^cA_{\nu}^d\lambdabar^a\lambdabar^b)\cr
&+id^{abc}gC^{\mu}{}_{\nu}A_{\mu}^a\lambdabar^b\sigmabar^{\nu\rho}\pa_{\rho}
\lambdabar^c-\frak12id^{cde}f^{abe}g^2C^{\mu\rho}A_{\mu}^cA_{\nu}^d
\lambdabar^a\sigmabar^{\nu\rho}\lambdabar^b\cr
&+iC^{\rho\sigma}d^{a0c}g_0A_{\sigma}^a\lambdabar^0[\delta^{\mu}{}_{\rho}
+2\sigmabar^{\mu}{}_{\rho}]\pa_{\mu}\lambdabar^c\cr
&+iC^{\mu\nu}f^{abc}d^{cd0}gg_0A_{\mu}^bA_{\nu}^d
\lambdabar^a\lambdabar^0\cr
&-\Bigl\{\frak12i\sqrt2g^2C^{\mu\nu}d^{abc}
A^b_{\mu}\phibar R^c\lambdabar^a\sigmabar_{\nu}\psi
+i\sqrt2gg_0C^{\mu\nu}d^{0bc}
A^b_{\mu}\phibar R^c\lambdabar^0\sigmabar_{\nu}\psi\cr
&+(\phi\go\phitil,\psi\go\psitil,F\go\Ftil,C^{\mu\nu}\go
-C^{\mu\nu})\Bigr\}
\Bigr],\cr}}
which miraculously casts all the $C$-dependent terms apart from those linear 
in $F$, $\Ftil$
 into the correct form. Then finally redefinitions of $\Fbar$, $\bar{\tilde F}$
 can be
used to deal with the terms linear in $F$, $\Ftil$. Explicitly, we 
need
\eqn\fbarredef{\eqalign{
\delta \Fbar=&
L\Bigl\{\Bigl[\left(5Ng^2+2(1+\alpha) \Chat_2\right)
g\pa_{\mu}A_{\nu}^a\cr
&-\left(\frak{11}{4}Ng^2+(1+\alpha)\Chat_2\right)
g^2f^{abc}A_{\mu}^bA_{\nu}^c\Bigr]iC^{\mu\nu}\phibar R^a\cr
&+2(\alpha+1)\Chat_2g_0\pa_{\mu}A_{\nu}^0iC^{\mu\nu}\phibar R^0\cr
&+\frak{1}{8}|C^2|\Bigl[\left(-37Ng^2+(63+\alpha)\Chat_2\right)
g^2d^{abc}\phibar R^c\lambdabar^a\lambdabar^b\cr
&+\left(-32Ng^2+(31+\alpha)\Chat_2\right)
gg_0d^{0bc}\phibar R^c\lambdabar^0\lambdabar^b\cr
&+(31+\alpha)\Chat_2
g_0^2d^{000}\phibar R^0\lambdabar^0\lambdabar^0
+\left(6Ng^2+(\alpha-1)\Chat_2\right)g^2
d^{ab0}\phibar R^0\lambdabar^a\lambdabar^b\Bigr]\Bigr\}\cr}}
(with a similar redefinition of $\bar{\tilde F}$)
which produce a change in the action
\eqn\Fchange{
\delta S_F=\int d^4x \left(\delta \Fbar F +\delta \bar{\tilde F} \Ftil\right).}
We now find that with
\eqn\zforms{
Z^{(1)}_C=Z_{|C|^2}^{(1)}=0,\quad Z^{(1)}_1=-3Ng^2L,\quad Z^{(1)}_2=-Ng^2L,}
we have
\eqn\gammone{
\Gamma^{(1)\rm{pole}\prime}=\sum_{i=1}^8\Gamma_{i\rm{1PI}}^{(1)\rm{pole}}
+\delta S_{\lambda}+\delta S_F+S_B^{(1)}=0,}
i.e. $\Gamma^{(1)\prime}$ is finite. 

This demonstrates that our theory is renormalisable and that the
$\Ncal=\frak12$ supersymmetry is preserved. However we find that to obtain a 
renormalisable lagrangian it is vital (since $Z^{(1)}_1, Z^{(1)}_2\ne0$)
to include the terms involving $\gamma_1$, $\gamma_2$  in Eq.~\lagranb,
which were not 
in the original formulation of the theory\araki\ though they are 
independently $\Ncal=\frak12$ 
supersymmetric. This is not unexpected since in general any terms which
are not forbidden by a symmetry will be generated under renormalisation.
It is therefore all the more remarkable that we do not need to renormalise 
the nonanticommutativity parameter $C$ and that the other $\lambdabar^4$ terms 
(which are also separately $\Ncal=\frak12$ supersymmetric) do not require any
counterterms. On the other hand our renormalised lagrangian is no longer of the
form derived from nonanticommutative superspace. Of course this was also found 
in the case of the $\Ncal=\frak12$ Wess-Zumino model\gris.  

We note here that the requirement to make a divergent redefinition of $\Fbar$
is not as surprising as it may first appear (if calculating in 
components with a conventional covariant gauge). In fact, if one renormalises
the ordinary $\Ncal=1$ theory in its uneliminated component form, i.e. before
eliminating the auxiliary fields, one is compelled to make a similar
non-linear renormalisation  of $F$ to render the theory finite. This has not
to our knowledge previously been discussed, and we give the details in 
a forthcoming publication\ref\jjwa{I.~Jack, D.R.T.~Jones
and L.A.~Worthy, in preparation}.   

\newsec{Conclusions}
We have studied the renormalisability of a general
$\Ncal=\frak12$ supersymmetric theory coupled to chiral matter.
The non-renormalisability of
the standard $U(N)$ version was apparent from the outset, and it appeared 
impossible to define a general 
$SU(N)$ invariant $\Ncal=\frak12$ supersymmetric theory;
however we were able to 
define an $SU(N)\times U(1)$ invariant action which still possessed
$\Ncal=\frak12$ supersymmetry, which as we showed was preserved
under renormalisation. Moreover we find that the non-anticommutativity
parameter $C$ is unrenormalised (at least at one loop).

We have restored gauge invariance by a somewhat unconventional 
expedient which works rather miraculously. One could speculate to 
what extent the $\Ncal=\frak12$ supersymmetry and the identities Eq.~\cprop{}\
were required to make this trick work. If one treats the action \lagrana\ 
as primordial, ignoring its derivation from non-anticommuting
superspace, the identities Eq.~\cprop{}\ can be
regarded as a consequence of the self-duality of $C^{\mu\nu}$ (with 
$C^{\alpha\beta}$ now {\it defined} by Eq.~\cprop{a}).    
It would be interesting to examine a theory of the same form but in which 
$C^{\mu\nu}$ was replaced by a general antisymmetric tensor. Moreover,
suppose one considered a theory with an action based on Eq.~\lagrana\ but
including all the hermitian conjugate terms which are missing. The only new 
diagrams would simply be the ``hermitian  conjugates'' of those in 
Figs.~1--8. Eq.~\lchange\ would now need to be supplemented by its hermitian
conjugate. However, the variation of the action would now include
additional unwanted non-gauge-invariant terms since it is now not only the 
gaugino kinetic term which varies. This raises the possibility of a theory 
(albeit non-renormalisable) with ineradicable non-gauge-invariant 
divergences.

An interesting feature of our results is the redefinition (or non-linear
renormalisation) of the gaugino field.
As we have mentioned, the attendant non-linear 
redefinition of the auxiliary field $F$ has its counterpart even in the $\Ncal=1$ 
theory, so that non-linear field redefinitions may be an unavoidable
consequence of working in the uneliminated component formalism with 
conventional gauge-fixing; as we
mentioned, no such field redefinition was required in the $\Ncal=\frak12$ 
superfield calculation of Ref.~\penrom.

\appendix{A}{Results for One-Loop Diagrams}
The divergent contributions 
to the effective action from the graphs in Fig.~1 are of the form:
\eqn\formone{\eqalign{
ig^2Ld^{ABC}e^{ABC}C^{\mu\nu}\Bigl[\pa_{\mu}A^A_{\rho}\lambdabar^B
\bigl(T_1^{ABC}\delta_{\nu}{}^{\rho}
&+\Atil^{ABC}_1\sigmabar_{\nu}{}^{\rho}\bigr)\lambdabar^C\cr
+A^A_{\rho}\lambdabar^B
\bigl(\Ttil_1^{ABC}\delta_{\nu}{}^{\rho}  
&+A^{ABC}_1\sigmabar_{\nu}{}^{\rho}\bigr)\pa_{\mu}\lambdabar^C\Bigr],\cr}}  
where the contributions to $T_1$, $\Ttil_1$, $A_1$, $\Atil_1$ from the 
individual graphs are given in Table 1:
\bigskip
\vbox{
\begintable
Graph|$T_1$|$\Ttil_1$|$\Atil_1$|$A_1$\cr
1a|$-\left(3+\alpha\right)Nd^Ac^Bc^C$|0|0|0\cr
1b|$-Nc^Ad^Bc^C$|0|$-\frak23Nc^Ac^Bd^C$|$-\frak43Nc^Ac^Bd^C$\cr
1c|$-2\alpha Nc^Ad^Bc^C$
|$\frak12(2-\alpha)Nc^Ad^Bc^C$|$\frak23Nc^Ad^Bc^C$|
$-\frak13(2+3\alpha)Nc^Ad^Bc^C$\cr
1d|$\frak12(5+\alpha)Nc^A$|0|0|0\cr
1e|0|$-\frak12(3-\alpha)Nc^Ad^Bc^C$|0|$(1+\alpha)Nc^Ad^Bc^C$\cr
1f|$-g^{ABC}/e^{ABC}$|0|0|$-\frak43g^{ABC}/e^{ABC}$\cr
1g|$-\frak12g^{ABC}/e^{ABC}$|0|0|
$-\frak23g^{ABC}/e^{ABC}$\cr
1h|$-\frak12g^{ABC}/e^{ABC}$|0|0|
$2g^{ABC}/e^{ABC}$
\endtable}
\centerline{{\it Table~1:\/} Contributions to $T_1$, $\Ttil_1$, 
$A_1$, $\Atil_1$ from Fig.~1}
\bigskip
In Table 1, $g^{a0b}=g^{ab0}=g^{0ab}=g^{000}=g_0$ and $g^{abc}=g$.
 
We note here that Figs.~1f-1h correspond to both $\phi$, $\psi$ and $\phitil$,
$\psitil$ loops, which contribute identically due to the change in 
sign $C^{\mu\nu}\go-C^{\mu\nu}$ between the $\phi$, $\psi$ and $\phitil$,
$\psitil$ interactions in the lagrangian. 
Possible contributions of the form
$gLf^{ABC}C^{\mu\nu}\pa_{\mu}A^A_{\rho}\lambdabar^B
\sigmabar_{\nu}{}^{\rho}\lambdabar^C$,
$gLf^{ABC}C^{\mu\nu}A^A_{\rho}\lambdabar^B
\delta_{\nu}{}^{\rho}\pa_{\mu}\lambdabar^C$
cancel between $\phi$, $\psi$ and $\phitil$,
$\psitil$ loops.

The divergent contributions
to the effective action from the graphs in Fig.~1 are given by
\eqn\sumone{\eqalign{
\Gamma_{1\rm{1PI}}^{(1)\rm{pole}}=&ig^2LC^{\mu\nu}d^{ABC}e^{ABC}
\Bigl[\pa_{\mu}A^A_{\rho}\lambdabar^B
\Bigl(\bigl[-(1+2\alpha)Nc^Ad^Bc^C+\frak12(5+\alpha)Nc^A
\cr
&-(3+\alpha)Nd^Ac^Bc^C-2{g^{ABC}\over{e^{ABC}}}\bigr]\delta_{\nu}{}^{\rho}
+\frak23N\bigl[c^Ad^Bc^C-c^Ac^Bd^C]
\sigmabar_{\nu}{}^{\rho}\Bigr)\lambdabar^C\cr
&+A^A_{\rho}\lambdabar^B
\bigl(-\frak12Nc^Ad^Bc^C\delta_{\nu}{}^{\rho}
+\frak13N\bigl[c^Ad^Bc^C-4c^Ac^Bd^C\bigr]
\sigmabar_{\nu}{}^{\rho}\bigr)\pa_{\mu}\lambdabar^C\Bigr]\cr
=&iLC^{\mu\nu}\Bigl[-\bigl\{\frak54(1+2\alpha)N+2\bigr\}g^3d^{abc}
\pa_{\mu}A^a_{\nu}\lambdabar^b\lambdabar^c\cr
&+[3(1-\alpha)N-4]g^2g_0d^{ab0}\pa_{\mu}A^a_{\nu}\lambdabar^b\lambdabar^0\cr
&-2[(3+\alpha)Ng^2+g_0^2]
{g^2\over{g_0}}d^{0bc}\pa_{\mu}A^0_{\nu}\lambdabar^b\lambdabar^c
-2g^2g_0d^{000}\pa_{\mu}A^0_{\nu}\lambdabar^0\lambdabar^0\cr
&-Ng^2g_0d^{a0c}A^a_{\nu}\lambdabar^0\pa_{\mu}\lambdabar^c\cr
&-Ng^3d^{abc}A^a_{\rho}\lambdabar^b\sigmabar_{\nu}{}^{\rho}\pa_{\mu}\lambdabar^c
-2Ng^2g_0d^{a0c}
A^a_{\rho}\lambdabar^0\sigmabar_{\nu}{}^{\rho}\pa_{\mu}\lambdabar^c\Bigl].\cr
}}
The divergent contributions 
to the effective action from the graphs in Fig.~2 are of the form:
\eqn\formtwo{\eqalign{
ig^3Le^{EAB}[&d^{ABE}f^{CDE}C^{\mu\nu}T_2^{ABCD}A_{\mu}^CA_{\nu}^D\lambdabar^A
\lambdabar^B\cr
&+d^{CDE}f^{ABE}C^{\mu\rho}A_2^{ABCD}A_{\mu}^CA_{\nu}^D\lambdabar^A
\sigmabar^{\nu}{}_{\rho}\lambdabar^B]\cr}}
where the contributions to $T_2$, $A_2$  from the
individual graphs are given in Table 2:  
\bigskip   
\vbox{  
\begintable
Graph|$T_2$|$A_2$\cr
2a|$\frak12Nd^Ac^B$|$\frak13Nd^Ac^Bc^Cc^D$\cr
2b|$-\frak12(3-\alpha)N\delta^{A0}$|\cr
2c|$\frak12(3+\alpha)Nc^Ac^B$|0\cr
2d|$\frak12(2-\alpha)N\delta^{A0}$|\cr
2e|$-\frak12\alpha Nd^Ac^B$|$\frak16(4+3\alpha)Nd^Ac^Bc^Cc^D$\cr
2f|$\frak34\alpha Nd^Ac^B$|$-\frak12(2+\alpha)Nd^Ac^Bc^Cc^D$\cr
2g|$\frak34\alpha Nd^Ac^B$|$\frak12 Nd^Ac^Bc^Cc^D$\cr
2h|$-\frak34(1+\alpha)N$|0\cr
2i|$\frak34\alpha N$|0\cr
2j|$\frak12$|$\frak13$\cr
2k|0|$\frak23$\cr
2l|$\frak12$|$-1$
\endtable}
\centerline{{\it Table~2:\/} Contributions from Fig.~2}
\bigskip

The contributions from Figs.~2(m)--(o) are zero. The graphs in Fig.~2 add to
\eqn\sumtwo{\eqalign{
\Gamma_{2\rm{1PI}}^{(1)\rm{pole}}
=&\frak14ig^3L\Bigl[2(1+2\alpha)Nd^Ac^B+2(3+\alpha)Nc^Ac^B-3N
-2N\delta^{A0}
+4\Bigr]\cr
&e^{EAB}d^{ABE}f^{CDE}C^{\mu\nu}A_{\mu}^CA_{\nu}^D\lambdabar^A
\lambdabar^B\cr
&+\frak12ig^3Ld^Ac^Bc^Cc^D
e^{EAB}d^{CDE}f^{ABE}
C^{\mu\rho}A_{\mu}^CA_{\nu}^D\lambdabar^A
\sigmabar^{\nu}{}_{\rho}\lambdabar^B]\cr
=&\left[\frak14(5+6\alpha)N+1\right]iL
d^{abe}f^{cde}g^4C^{\mu\nu}A_{\mu}^cA_{\nu}^d\lambdabar^a\lambdabar^b\cr
&+\frak12iNLd^{cde}f^{abe}g^4C^{\mu\rho}A_{\mu}^cA_{\nu}^d
\lambdabar^a\sigmabar^{\nu}{}_{\rho}\lambdabar^b\cr
&-iL[(1-2\alpha)N-2]
d^{0be}f^{cde}g^3g_0C^{\mu\nu}A_{\mu}^cA_{\nu}^d\lambdabar^0\lambdabar^b.\cr
}}
The results
for Fig.~3 are of the form: 
\eqn\formthree{
g^2L|C|^2\left[X_1^{abcd}
(\lambdabar^a\lambdabar^b)(\lambdabar^c\lambdabar^d)
+X_2(\lambdabar^a\lambdabar^a)(\lambdabar^b\lambdabar^b)
+X_3(\lambdabar^a\lambdabar^a)(\lambdabar^0\lambdabar^0)]\right]}
where the contributions to $X_{1-3}$ are given in Table 3:
\bigskip
\vbox{   
\begintable
Graph|$X_1^{abcd}$|$X_2$|$X_3$\cr
3a|$\frak14(3+\alpha)Ng^2d^{abe}d^{cde}+2g^2d^{abcd}+(1-\alpha)g^2d^{adcb}
-\frak4N{g^4\over{g_0^2}}f^{eac}f^{ebd}$|$(3+\alpha){g^4\over{g_0^2}}$|
$0$\cr
3b|$\frak12\alpha Ng^2d^{abe}d^{cde}$|0|$-2\alpha g_0^2$\cr
3c|$-\frak14(1+\alpha)Ng^2d^{abe}d^{cde}$|0|$(1+\alpha) g_0^2$\cr
3d|$g^2\left[-2d^{abcd}+(\alpha-1)d^{adcb}+\frak4N{g^2\over{g_0^2}}
f^{eac}f^{ebd}\right]$|0|
$(3+\alpha)g_0^2$\cr
3e|$\frak13g^2(\dtil^{abcd}-\dtil^{acdb})$|0|$-g_0^2$\cr
3f|$\frak14g^2d^{abe}d^{cde}$|$\frak{1}{2N}g^2$|0
\endtable}
\centerline{{\it Table~3:\/} Contributions from Fig.~3}
\bigskip
In Table 3,
\eqn\dabcddefs{
d^{abcd}=\Tr[F^aF^bD^cD^d], \qquad \dtil^{abcd}=\Tr[F^aD^cF^bD^d],}
where the matrices $F_a$ and $D_a$ are defined in Appendix B.
These results add to
\eqn\sumthree{\eqalign{
\Gamma_{3\rm{1PI}}^{(1)\rm{pole}}=&g^2L|C|^2\Bigl[
\frak14[(3+2\alpha)N+1]g^2d^{abe}d^{cde}(\lambdabar^a\lambdabar^b)
(\lambdabar^c\lambdabar^d)\cr
&+\frak{1}{2N}\left[2(3+\alpha)N\frak{g^4}{g_0^2}+g^2\right]
(\lambdabar^a\lambdabar^a)
(\lambdabar^b\lambdabar^b)+3g_0^2(\lambdabar^a\lambdabar^a)
(\lambdabar^0\lambdabar^0)\Bigr].\cr}}
In obtaining these results we have made frequent use of the 
Fierz identity
\eqn\fierz{
(\lambdabar^a\lambdabar^b)(\lambdabar^c\lambdabar^d)
+(\lambdabar^a\lambdabar^c)(\lambdabar^b\lambdabar^d)
+(\lambdabar^a\lambdabar^d)(\lambdabar^b\lambdabar^c)=0}

The contributions from the graphs shown in Fig.~4 are of the form
\eqn\formfour{
\sqrt2g_ALC^{\mu\nu}\pa_{\mu}\phibar \lambdabar^A X^A\sigmabar_{\nu}\psi
+\sqrt2g_ALC^{\mu\nu}\phibar \lambdabar^A Y^A\sigmabar_{\nu}\pa_{\mu}\psi  }
where $g_a\equiv g$ and 
$X^A$ and $Y^A$ are as given in Table~4. (There are analogous diagrams
with $\phitil$, $\psitil$ external legs which we do not show explicitly;
their contributions may easily be read off using
$\phi\go\phitil$, $\psi\go\psitil$, $F\go\Ftil$, $C^{\mu\nu}\go
-C^{\mu\nu}$.)   The contributions to $X^A$, $Y^A$ are shown in Table 4:
\bigskip
\vbox{
\begintable
Graph| $X^A$| $Y^A$\cr
4a|$\frak32Ng^2c^AR^A$|$-\alpha Nc^AR^A$\cr
4b|$\alpha Ng^2c^AR^A$|$\alpha Nc^AR^A$\cr
4c|$\alpha Ng^2c^AR^A$|$0$\cr
4d|$-2[\Chat_2-\frak12 Ng^2c^A]R^A$|$2[\Chat_2-\frak12 Ng^2c^A]R^A$\cr
4e|$-2[\Chat_2-\frak12 Ng^2c^A]R^A$|$0$\cr
4f|$-(1-2\alpha)[\Chat_2-\frak12 Ng^2c^A]R^A$|$0$\cr
4g|$2[2\Chat_2-\frak12 Ng^2c^A]R^A$|$0$\cr
4h|$2[2\Chat_2-\frak12 Ng^2c^A]R^A$|$-2[2\Chat_2-\frak12 Ng^2c^A]R^A$\cr
4i|$0$|$2[\Chat_2-\frak12 Ng^2c^A]R^A$\cr
4j|$-3\Chat_2R^A$|$0$   
\endtable}
\centerline{{\it Table~4:\/} Contributions to $X^A$ and $Y^A$ from Fig.~4}
\bigskip

These graphs add to
\eqn\sumfour{\eqalign{
\Gamma_{4\rm{1PI}}^{(1)\rm{pole}}
=&\sqrt2g_ALC^{\mu\nu}\pa_{\mu}\phibar \lambdabar^A
\Bigl[2\alpha\Chat_2R^A+(2+\alpha)Ng^2c^AR^A\Bigr]
\sigmabar_{\nu}\psi\cr
&-\sqrt2Ng_Ag^2c^AC^{\mu\nu}\phibar \lambdabar^A
R^A
\sigmabar_{\nu}\pa_{\mu}\psi\cr
=&L\Bigl\{\left[2\alpha\Chat_2+(2+\alpha)Ng^2\right]
\sqrt2gC^{\mu\nu}\pa_{\mu}\phibar \lambdabar^aR^a\sigmabar_{\nu}\psi
\cr
&+2\alpha\Chat_2
\sqrt2g_0C^{\mu\nu}\pa_{\mu}\phibar \lambdabar^0R^0\sigmabar_{\nu}\psi\cr
&-N
\sqrt2g^3C^{\mu\nu}\phibar \lambdabar^aR^a\sigmabar_{\nu}\pa_{\mu}\psi\Bigr\}.
\cr}  }
The contributions from the graphs shown in Fig.~5 are of the form
\eqn\formfive{
\sqrt2i
g_Ag_BLC^{\mu\nu}A_{\mu}^B\phibar \lambdabar^A Z^{AB}\sigmabar_{\nu}\psi}
where in the case of Figs. 5(a)--5(v), $Z^{AB}$ contains the contributions 
shown in Table~5:
\bigskip
\vbox{
\begintable
Graph|$\Chat_2R^AR^B$|$\Chat_2R^BR^A$|$Ng^2R^AR^B$|$Ng^2R^BR^A$|
$g^2f^{ACE}f^{BDE}R^CR^D$\cr
5a|$0$|$2$|$-c^A$|$-c^B$|$2c^Ac^B$\cr
5b|$-2$|$0$|$c^A+c^B$|$0$|$-2c^Ac^B$\cr
5c|$0$|$0$|$2c^B$|$0$|$-4c^Ac^B$\cr
5d|$4$|$0$|$-2(c^A+c^B)$|$0$|$4c^Ac^B$\cr
5e|$0$|$2$|$0$|$-c^A$|$0$\cr
5f|$0$|$0$|$\frak12(1-\alpha)c^B$|$0$|$(\alpha-1)c^Ac^B$\cr
5g|2|0|$-(c^A+c^B)$|0|$2c^Ac^B$\cr
5h|0|$-\alpha$|0|$\frak12\alpha c^B$|0\cr
5i|0|0|0|$-\frak34\alpha c^B$|0\cr
5j|0|$3+\alpha$|0|$-\frak14(3+\alpha)c^B$|0\cr
5k|0|0|$-\frak14\alpha c^B$|0|$\frak12\alpha c^Ac^B$\cr
5l|0|$1-\alpha$|$\frak14(\alpha-1)c^A$|$\frak14(\alpha-1)(c^A+c^B)$|
$\frak12(1-\alpha)c^Ac^B$\cr
5m|0|$-2\alpha$|$\alpha c^A$|$\alpha c^B$|$-2\alpha c^Ac^B$\cr
5n|0|0|$-\alpha c^A$|0|$2\alpha c^Ac^B$\cr
5o|0|$\alpha$|$-\frak12\alpha c^A$|$-\frak12\alpha c^B$|$\alpha c^Ac^B$\cr
5p|0|0|$-\frak14(3+\alpha)c^A$|$-\frak14(3+\alpha)c^A$|
$\frak12(3+\alpha)c^Ac^B$\cr
5q|0|0|0|$-\alpha c^A$|0\cr
5r|0|0|$\frak12 \alpha c^A$|0|$-\alpha c^Ac^B$\cr
5s|0|0|$\frak32(1+\alpha)c^B$|$-\frak32(1+\alpha)c^B$|$-3(1+\alpha)c^Ac^B$\cr
5t|0|0|0|0|$-2\alpha c^Ac^B$\cr
5u|0|0|0|0|$2\alpha c^Ac^B$\cr
5v|0|0|$-\frak34\alpha c^B$|$\frak34\alpha c^B$|$\frak32\alpha c^Ac^B$
\endtable} 
\centerline{{\it Table~5a:\/} Contributions to $Z^{AB}$ from Figs.~5(a)--5(v)}
\bigskip
The contributions from Table 5a add to
\eqn\fivepart{\eqalign{
&i\sqrt2g_Ag_BLC^{\mu\nu}A_{\mu}^B\phibar \lambdabar^A 
\Bigl[4\Chat_2R^AR^B+(8-2\alpha)\Chat_2R^BR^A\cr
&+Ng^2(-4R^AR^Bc^A+2R^AR^Bc^B
-(2+\alpha)R^BR^Ac^A-\frak12(7+\alpha)R^BR^Ac^B)\Bigr]\sigmabar_{\nu}\psi.\cr}}

In the case of Figs. 5(w)--5(cc), the contributions to $Z^{ab}$ are shown in
Table 5b:
\bigskip
\vbox{
\begintable
Graph|$Ng^2R^aR^b$|$Ng^2R^bR^a$|$\frak1N\Delta R^aR^b$|
$\frak1N\Delta R^bR^a$|$g^2\delta^{ab}$\cr
5w|$-\frak12(3+\alpha)$|0|$-(3+\alpha)$|$3+\alpha$|$\frak14(3+\alpha)$\cr
5x|0|$-1$|0|0|$\frak12$\cr
5y|0|0|0|$-4$|$-1$\cr
5z|$\frak12(2+\alpha)$|0|$2+\alpha$|$-(2+\alpha)$|$-\frak14(2+\alpha)$\cr
5aa|$-\frak12\alpha$|0|$\alpha$|$-\alpha$|$\frak14\alpha$\cr
5bb|0|$-\frak12$|$-1$|$-1$|$-\frak14$\cr
5cc|$\frak12\alpha$|0|$-\alpha$|$\alpha$|$-\frak14\alpha$
\endtable}
\centerline{{\it Table~5b:\/} Contributions to $Z^{ab}$ from Fig.~5(w)--5(cc)}
\bigskip
The contributions to $Z^{0b}$ from Figs. 5(w)--5(cc) are shown in
Table 5c:
\bigskip
\vbox{
\begintable
Graph|$Ng^2R^bR^0$|$\frak1N\Delta R^0R^b$\cr
5w|$-3+\alpha$|0\cr
5x|$-2$|0\cr
5y|0|$-4$\cr
5z|$2+\alpha$|0\cr
5aa|$-\alpha$|0\cr
5bb|$-1$|$-2$\cr
5cc|$\alpha$|0
\endtable}
\centerline{{\it Table~5c:\/} Contributions to $Z^{0b}$ from Fig.~5(w)--5(cc)}
\bigskip
\vbox{
\begintable
Graph|$(a0)$|$(00)$\cr
|$g^2R^aR^0+2\frak1N\Delta R^aR^0$|
$g^2+\frak{1}{N^2}\Delta$\cr
5y|$-2$|$-4$\cr
5bb|$-1$|$-2$
\endtable}
\centerline{{\it Table~5d:\/} Contributions to $Z^{a0}$ and $Z^{00}$
from Fig.~5(y), 5(bb)}
\bigskip
The contributions to $Z^{a0}$ and $Z^{00}$
from Figs. 5(w)--5(cc) are shown in
Table 5d (those not shown explicitly are zero).
Adding the results from Table 5a in Eq.~\fivepart\ to those from 
Tables 5b--5d, we obtain
\eqn\sumfive{\eqalign{
\Gamma_{5\rm{1PI}}^{(1)\rm{pole}}
=&i\sqrt2NLC^{\mu\nu}\Bigl[g^4A_{\mu}^b\phibar \lambdabar^a
\Bigl[\frak12d^{abc}R^c
-R^aR^b-\frak12(7+3\alpha)R^bR^a\Bigl]
\sigmabar_{\nu}\psi\cr
&+g^3g_0\left[d^{0bc}R^c-\frak12(3+\alpha)\right]
A_{\mu}^b\phibar \lambdabar^0
R^0R^b\sigmabar_{\nu}\psi\cr
&-(3+\alpha)g^3g_0A_{\mu}^0\phibar \lambdabar^a
R^aR^0\sigmabar_{\nu}\psi\cr
&-2{\alpha\over {N^2}}\Chat_2
A_{\mu}^A\phibar\lambdabar^B 
g_Ag_BR^BR^A\sigmabar_{\nu}\psi\Bigr].\cr}}

The contributions from Fig.~6 are of the form
\eqn\formsix{
iLC^{\mu\nu}(g_A\pa_{\mu}A^A_{\nu}
\phibar XR^AF+g_AA^A_{\nu}\pa_{\mu}\phibar YR^AF)}
where $X$ and $Y$ are given in Table 6:
\bigskip
\vbox{
\begintable
Graph|$X$|$Y$\cr
6a|0|$3Ng^2c^A$\cr
6b|0|$2[2\Chat_2-Ng^2c^A]$\cr
6c|$-[4\Chat_2-Ng^2c^A]$|$-[4\Chat_2-Ng^2c^A]$\cr
6d|$-(5+\alpha)Ng^2c^A$|0\cr
6e|$2\alpha Ng^2c^A$|$-2Ng^2c^A$
\endtable}
\centerline{{\it Table~6:\/} Contributions from Fig.~6}
\bigskip

The contributions in Table~6 add to
\eqn\sumsix{\eqalign{
\Gamma_{6\rm{1PI}}^{(1)\rm{pole}}=&-iLC^{\mu\nu}\phibar g_A\pa_{\mu}A^A_{\nu}
\left[4\Chat_2+\left(4-
\alpha\right)Ng^2c^A\right]R^AF\cr
=&-iLC^{\mu\nu}\phibar\left\{g\left[4\Chat_2+\left(4-
\alpha\right)Ng^2\right]
\pa_{\mu}A^a_{\nu}R^a
+4g_0\Chat_2\pa_{\mu}A^0_{\nu}R^0\right\}F.
\cr}}
The contributions from Fig.~7 are of the form
\eqn\formseven{
ig^2LC^{\mu\nu}A^a_{\mu}A^b_{\nu}\phibar Zf^{abc}R^cF}
where $Z$ is given in Table 7:
\bigskip
\vbox{
\begintable
Graph|$Z$\cr
7a|$-\frak34\alpha Ng^2$\cr
7b|0\cr
7c|0\cr
7d|0\cr
7e|$-\frak14(2+\alpha)Ng^2$\cr
7f|$2\Chat_2-Ng^2$\cr
7g|$-\frak32\alpha Ng^2$\cr
7h|$\frak32(1+\alpha)Ng^2$\cr
7i|$\frak14(3+\alpha) Ng^2$\cr
7j|$\frak12\alpha Ng^2$\cr
7k|$-\frak34\alpha Ng^2$\cr
7l|0
\endtable}
\centerline{{\it Table~7:\/} Contributions from Fig.~7}
\bigskip
\vbox{
\begintable
Graph|$ab$|$a0$|$00$\cr  
8a|0|0|0\cr
8b|$-g^2\delta^{ab}-\frak4N\Delta R^aR^b$|$-\sqrt{\frak N2}
\left[g^2+
\frak{4}{N^2}\Delta\right]R^a$
|$-2g^2-\frak{2}{N^2}\Delta$\cr
8c|$\frak12g^2\delta^{ab}+\frak1N\Delta R^aR^b$
|$\frak14\left(\frak{2}{N}\right)^{\frak32}\Delta R^a$|
$\frak12g^2+\frak{1}{2N^2}\Delta$\cr
8d|$-\alpha g^2Nd^{abc}R^c$|$-\alpha g^2\sqrt{2N}$|0\cr
8e|$(1+\alpha)g^2Nd^{abc}R^c$|$(1+\alpha)g^2\sqrt{2N}$|0\cr
8f|$-\frak12\alpha g^2Nd^{abc}R^c$|$-\frak12\alpha g^2\sqrt{2N}$|0\cr
8g|0|0|0\cr
8h|$\frak12\alpha g^2Nd^{abc}R^c$|$\frak12\alpha g^2\sqrt{2N}R^a$|0\cr
8i|$\frak14(3+\alpha)g^2[\frak12Nd^{abc}R^c+\delta^{ab}]$|0|0\cr
8j|$\frak18\alpha g^2Nd^{abc}R^c$|$\frak18\alpha g^2\sqrt{2N}R^a$|0\cr
8k|$-\frak14g^2\delta^{ab}-\frak1N\Delta R^aR^b$|
$-\frak14\sqrt{\frak N2}
\left[g^2+
\frak{4}{N^2}\Delta\right]R^a$
|$-\frak12g^2-\frak{1}{2N^2}\Delta$
\endtable}
\centerline{{\it Table~8:\/} Contributions from Fig.~8}
\bigskip

The contributions in Table~7 add to
\eqn\sumseven{
\Gamma_{7\rm{1PI}}^{(1)\rm{pole}}=ig^2L
C^{\mu\nu}A^a_{\mu}A^b_{\nu}\phibar\left[2\Chat_2+\frak14(3-4\alpha)Ng^2
\right]f^{abc}R^cF}
The contributions from Fig.~8 are of the form
\eqn\formeight{
Lg_Ag_B|C|^2\lambdabar^A\lambdabar^B\phibar Z^{AB}F}
where the contributions to $Z$ are given in Table 8.
The contributions in Table~8 add to
\eqn\sumeight{\eqalign{
\Gamma_{8\rm{1PI}}^{(1)\rm{pole}}=
&L|C|^2\phibar\bigl\{g^2\left[\frak18(43+2\alpha)Ng^2-8\Chat_2\right]
\lambdabar^a\lambdabar^bd^{abc}R^c\cr
&+gg_0\left[\frak14(19+\alpha)Ng^2-8\Chat_2\right]
\lambdabar^0\lambdabar^bd^{0bc}R^c\cr
&+\frak14\alpha Ng^4d^{0bc}\lambdabar^b\lambdabar^c
-4Ng_0^2\Chat_2d^{000}\lambdabar^0\lambdabar^0R^0\bigr\}F.
\cr}}

\appendix{B} {Group identities for $SU(N)$}
The basic
commutation relations for $SU(N)$ are (for the fundamental representation): 
\eqn\commrel{ [R^a,R^b]=if^{abc}R^c,\quad
\{R^a,R^b\}=d^{abc}R^c+{1\over N}\delta^{ab},} 
where $d^{abc}$ is totally symmetric.
Defining matrices $F^a$, $D^a$ by $(F^a)^{bc}=f^{bac}$, $(D^a)^{bc}=d^{bac}$,
useful identities for $SU(N)$ are
\eqn\sunidents{\eqalign{
\Tr[F^aF^b]=&-N\delta^{ab},\qquad \Tr[D^aD^b]={N^2-4\over{N}}\delta^{ab},\cr
\Tr[F^aF^bD^c]=&-\frak{N}{2}d^{abc},\qquad 
\Tr[F^aD^bD^c]={N^2-4\over{2N}}f^{abc},\cr
&C_2(R)={N^2-1\over{2N}},\cr
\Tr[F^aD^bF^cD^d]=&
\frak{N}{4}(d^{acx}d^{bdx}-d^{abx}d^{cdx}-d^{adx}d^{bcx}).\cr}}
\eqn\sunidentsb{\eqalign{
d^{acd}R^bR^cR^d=&{N^2-4\over{2N}}R^bR^a,\cr
d^{ace}f^{bde}R^cR^d=&i\left[-\frak12NR^aR^b+\frak1N[R^a,R^b]
+\frak14\delta^{ab}\right],\cr
d^{ace}f^{bde}R^dR^c=&i\left[\frak12NR^bR^a+\frak1N[R^a,R^b]
-\frak14\delta^{ab}\right],\cr
d^{acd}R^cR^bR^d=&-\frak1N\{R^a,R^b\}+\frak14\delta^{ab},\cr}}

\bigskip
\epsfysize= 4in
\centerline{\epsfbox{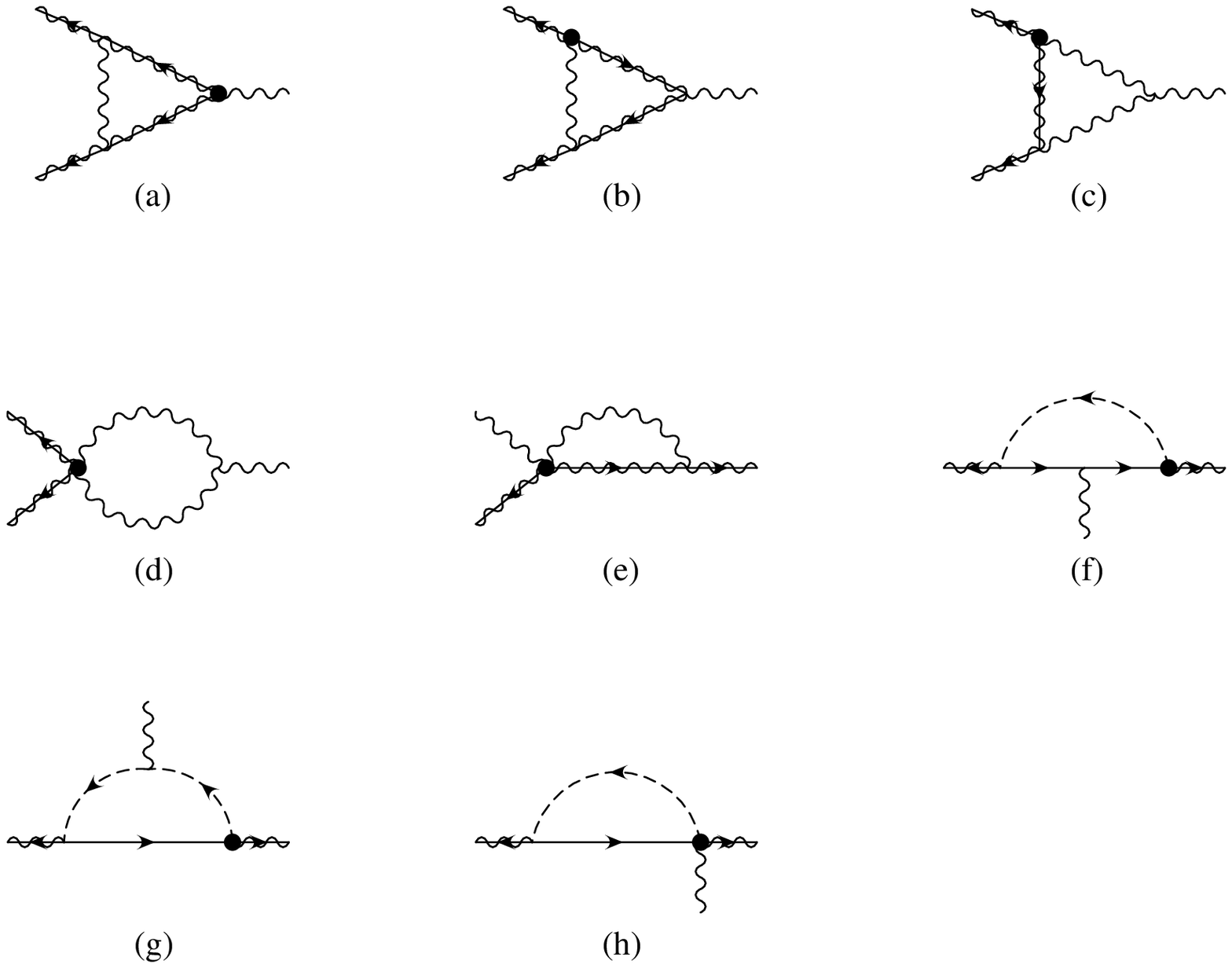}}
\inparg
{\it \noindent Fig. 1: Diagrams with one gauge, two gaugino lines; the dot
represents the position of a $C$.}
\medskip
\outparg
\vfill
\eject
\bigskip
\epsfysize= 7in
\centerline{\epsfbox{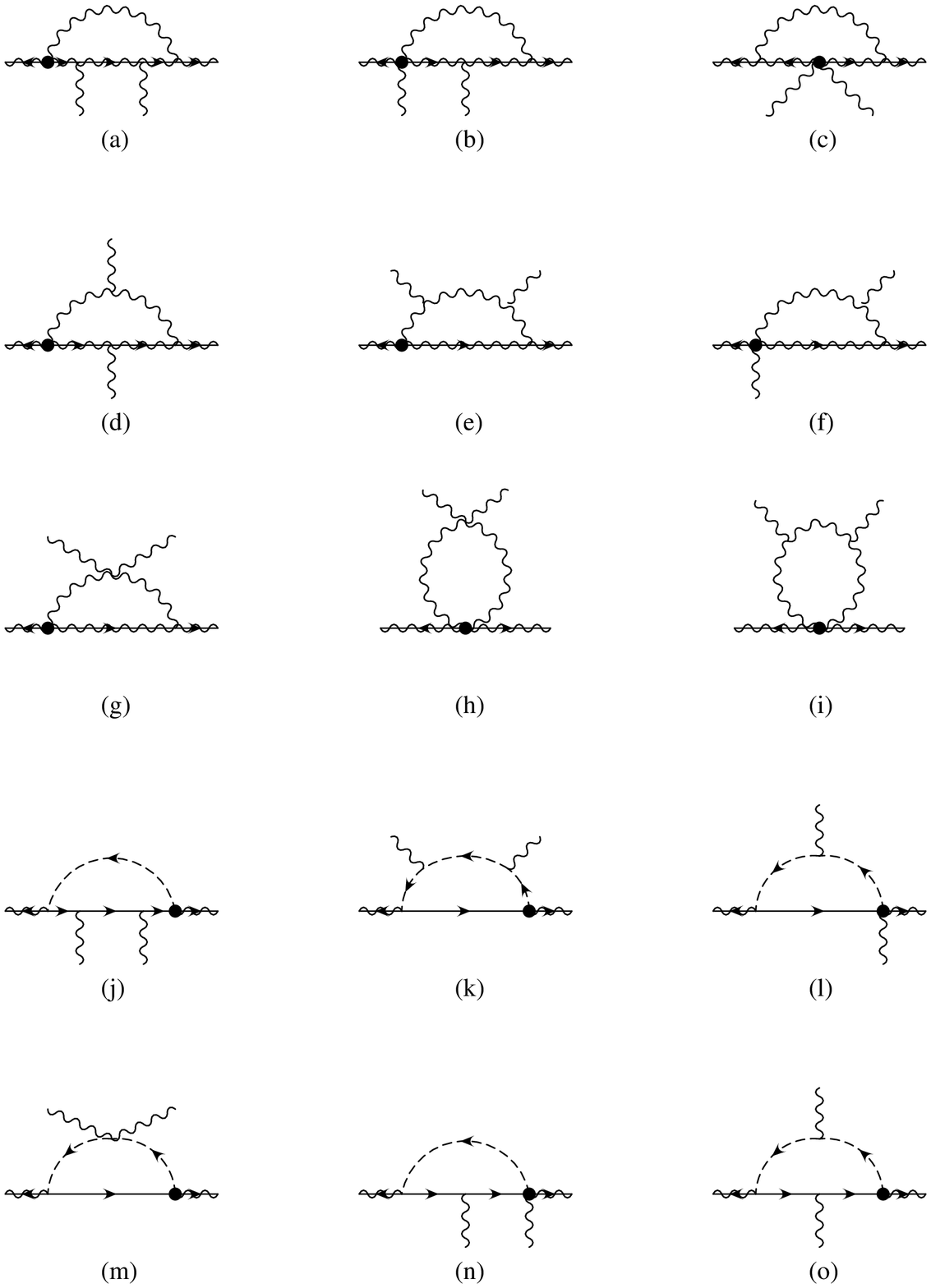}}
\inparg
{\it \noindent Fig. 2: Diagrams with two gauge and two gaugino lines; the 
dot represents the position of a $C$.}
\medskip
\outparg

\bigskip
\epsfysize= 2in
\centerline{\epsfbox{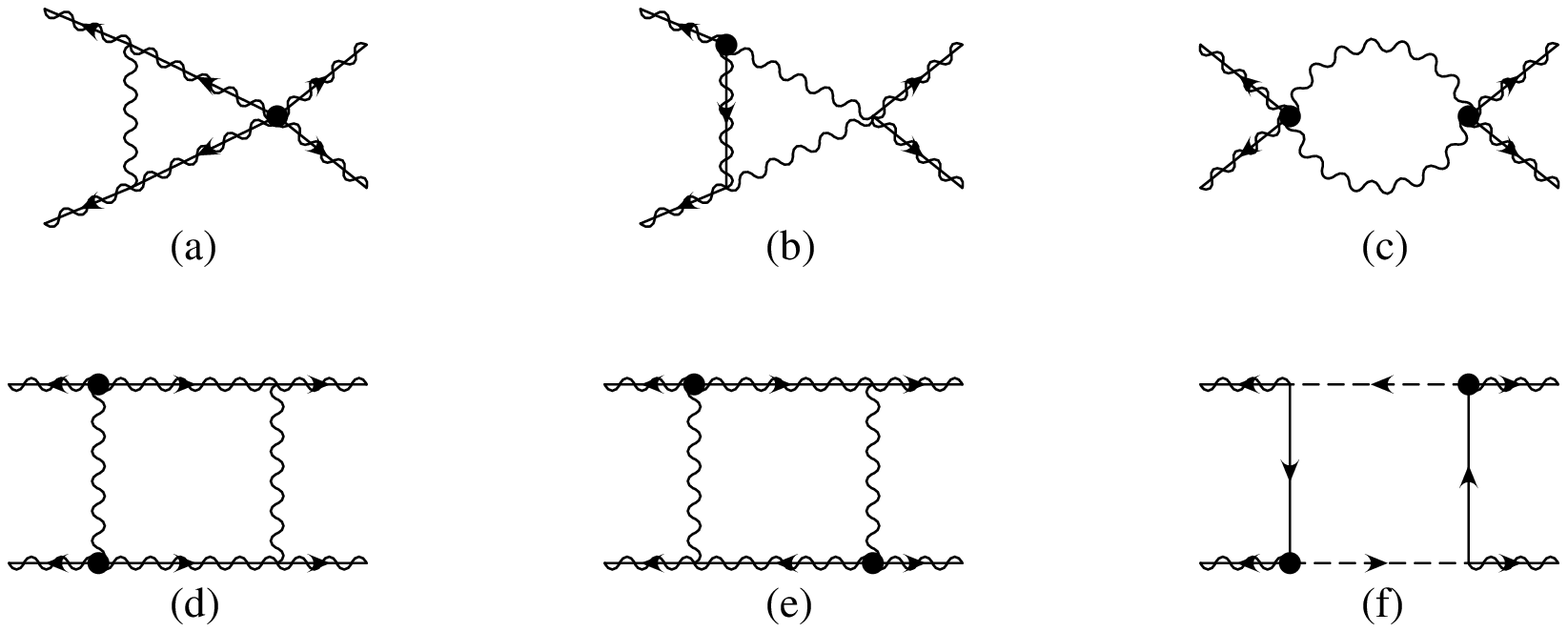}}
\inparg
{\it \noindent Fig. 3: Diagrams with four gaugino lines; the dot 
represents the position of a $C$ or $|C|^2$.}
\medskip
\outparg
\vfill
\eject
\bigskip
\epsfysize= 5in
\centerline{\epsfbox{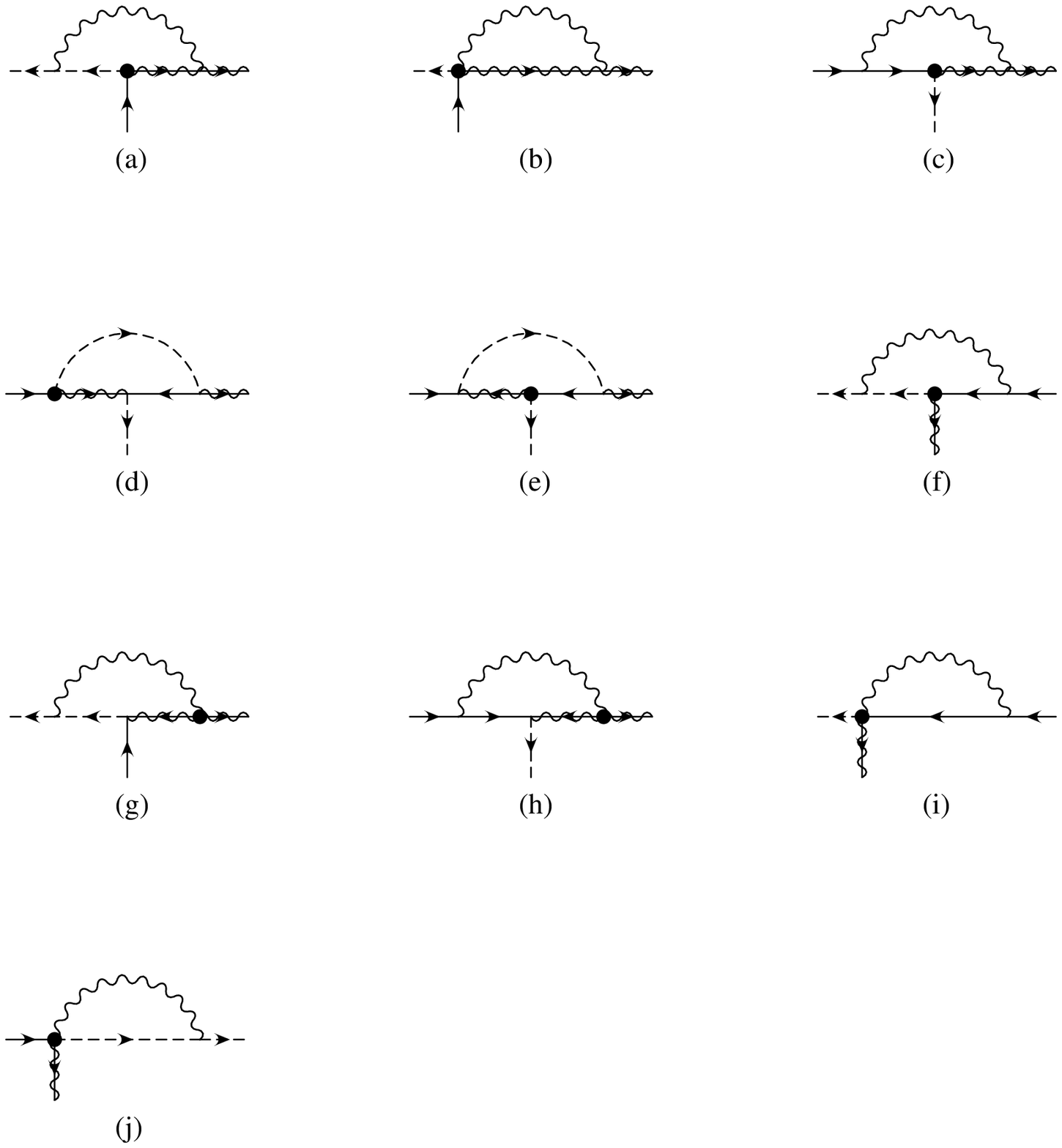}}
\inparg
{\it \noindent Fig. 4: Diagrams with one gaugino, one scalar and one
chiral fermion line; the dot represents the position of a $C$.}
\medskip
\outparg
\vfill
\eject
\bigskip
\epsfysize= 7in
\centerline{\epsfbox{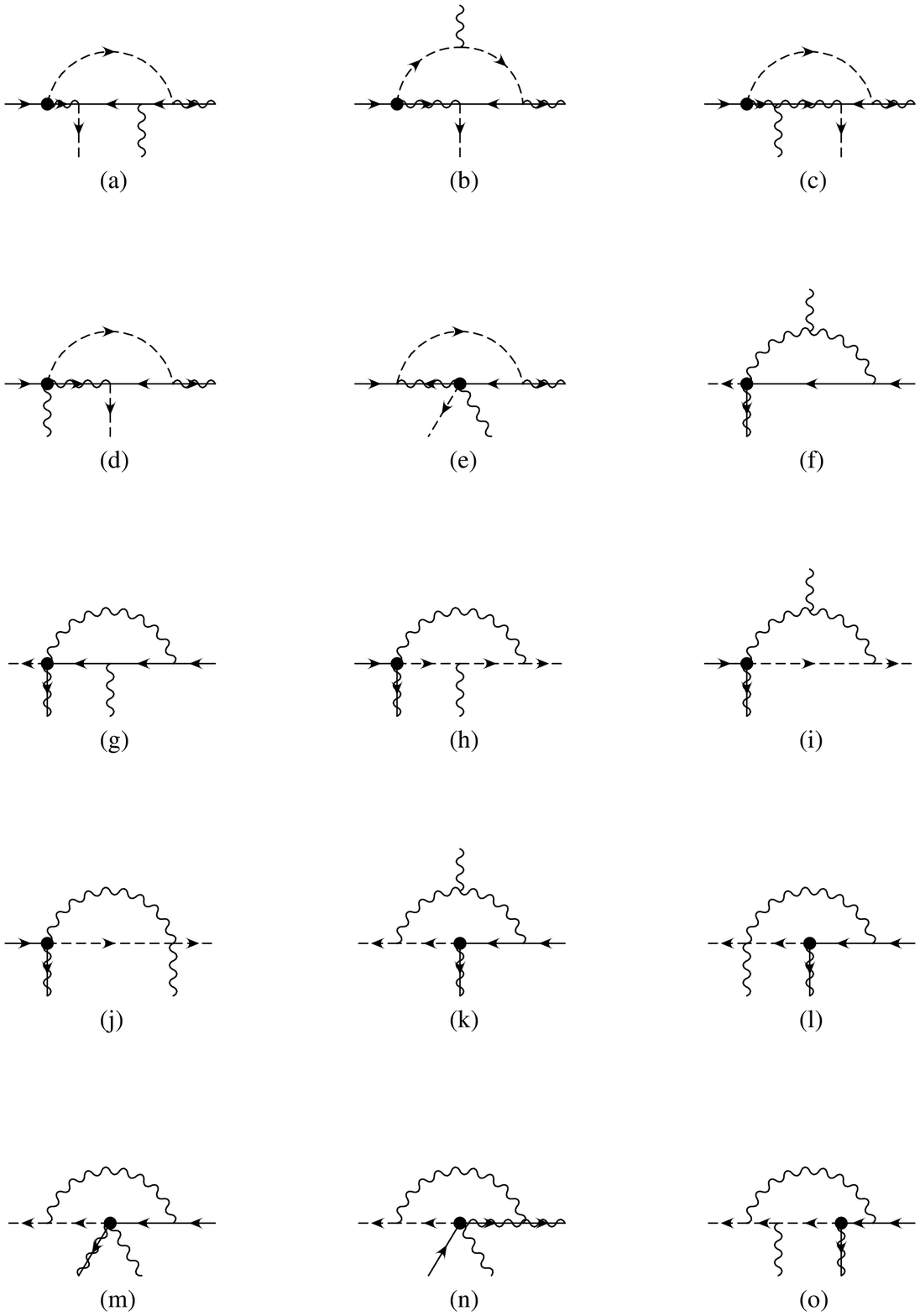}}
\inparg
{\it \noindent Fig. 5: Diagrams with one gaugino, one scalar, one
chiral fermion and one gauge line; the dot represents the position of a $C$.}
\medskip
\outparg

\bigskip
\epsfysize= 4in
\centerline{\epsfbox{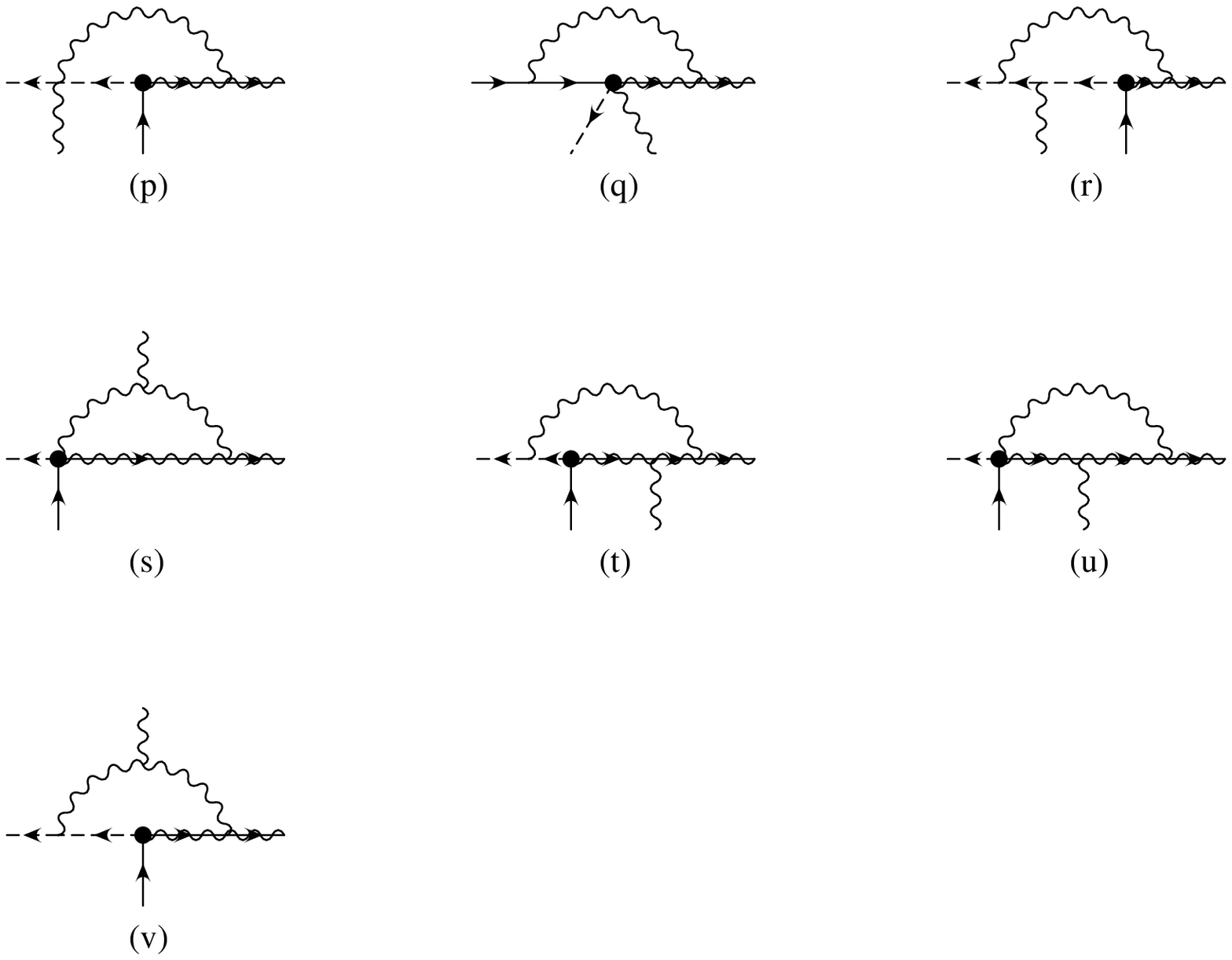}}
\inparg 
{\it \noindent Fig. 5(ctd): Diagrams with one gaugino, one scalar, one
chiral fermion and one gauge line; the dot represents the position of a $C$.}
\medskip
\outparg
\vfill
\eject
\bigskip
\epsfysize= 4in
\centerline{\epsfbox{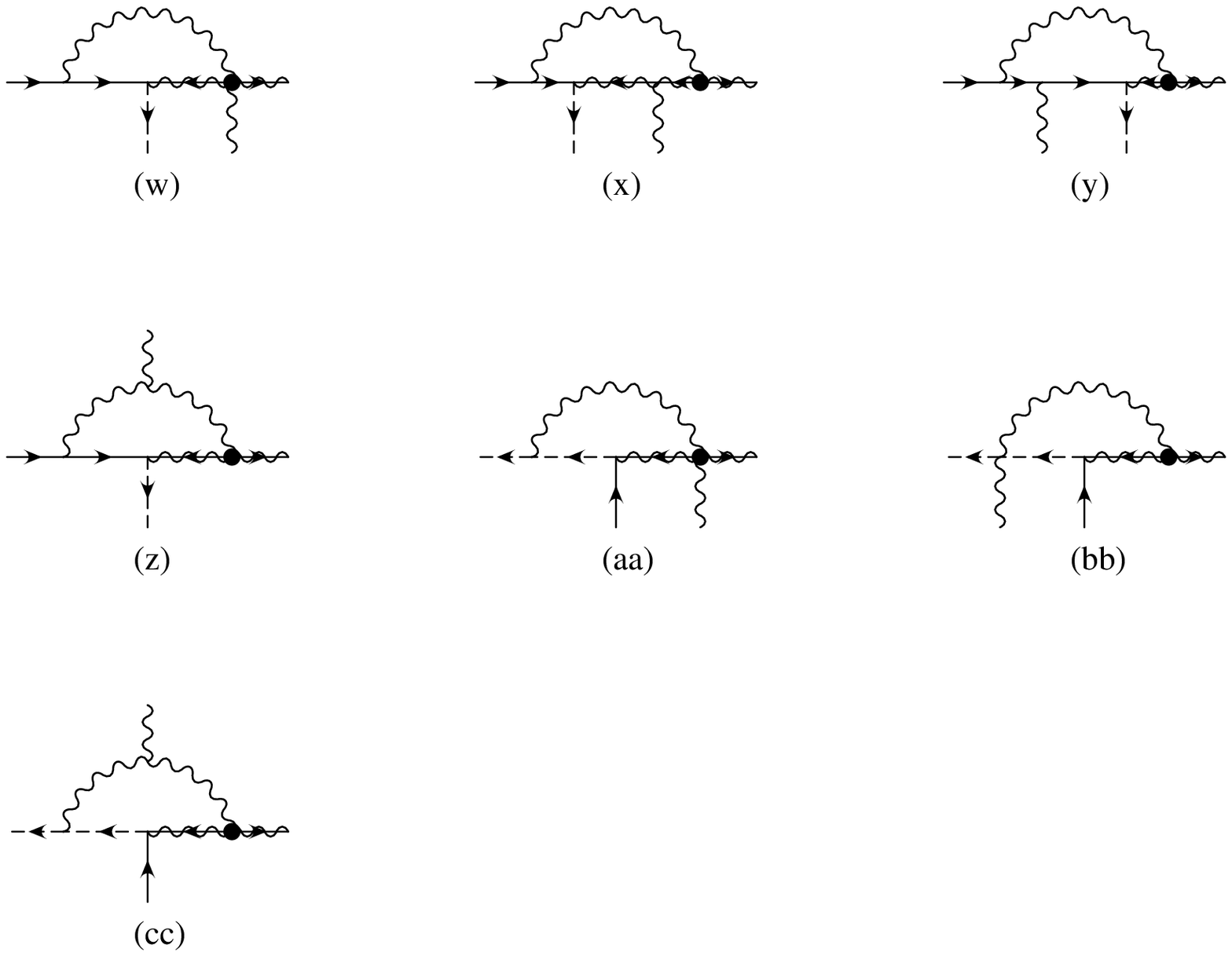}}
\inparg 
{\it \noindent Fig. 5b: Diagrams with one gaugino, one scalar, one
chiral fermion and one gauge line contributing an explicit $d^{abc}$; 
the dot represents the position of a $C$.}
\medskip
\outparg

\bigskip
\epsfysize= 2in
\centerline{\epsfbox{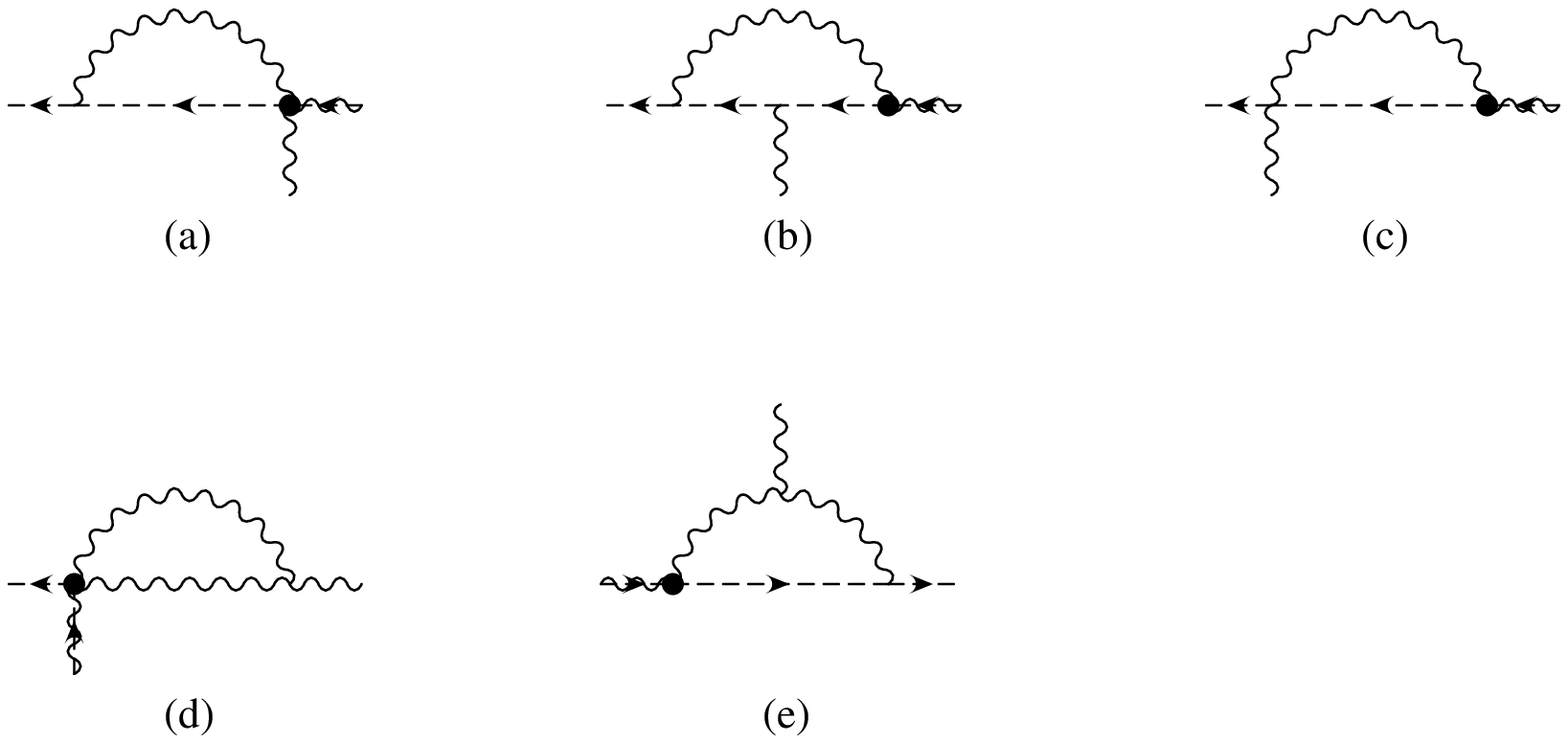}}
\inparg
{\it \noindent Fig. 6: Diagrams with one gauge, one scalar and one 
auxiliary line; the dot represents the position of a $C$.}
\medskip
\outparg

\bigskip
\epsfysize= 5in
\centerline{\epsfbox{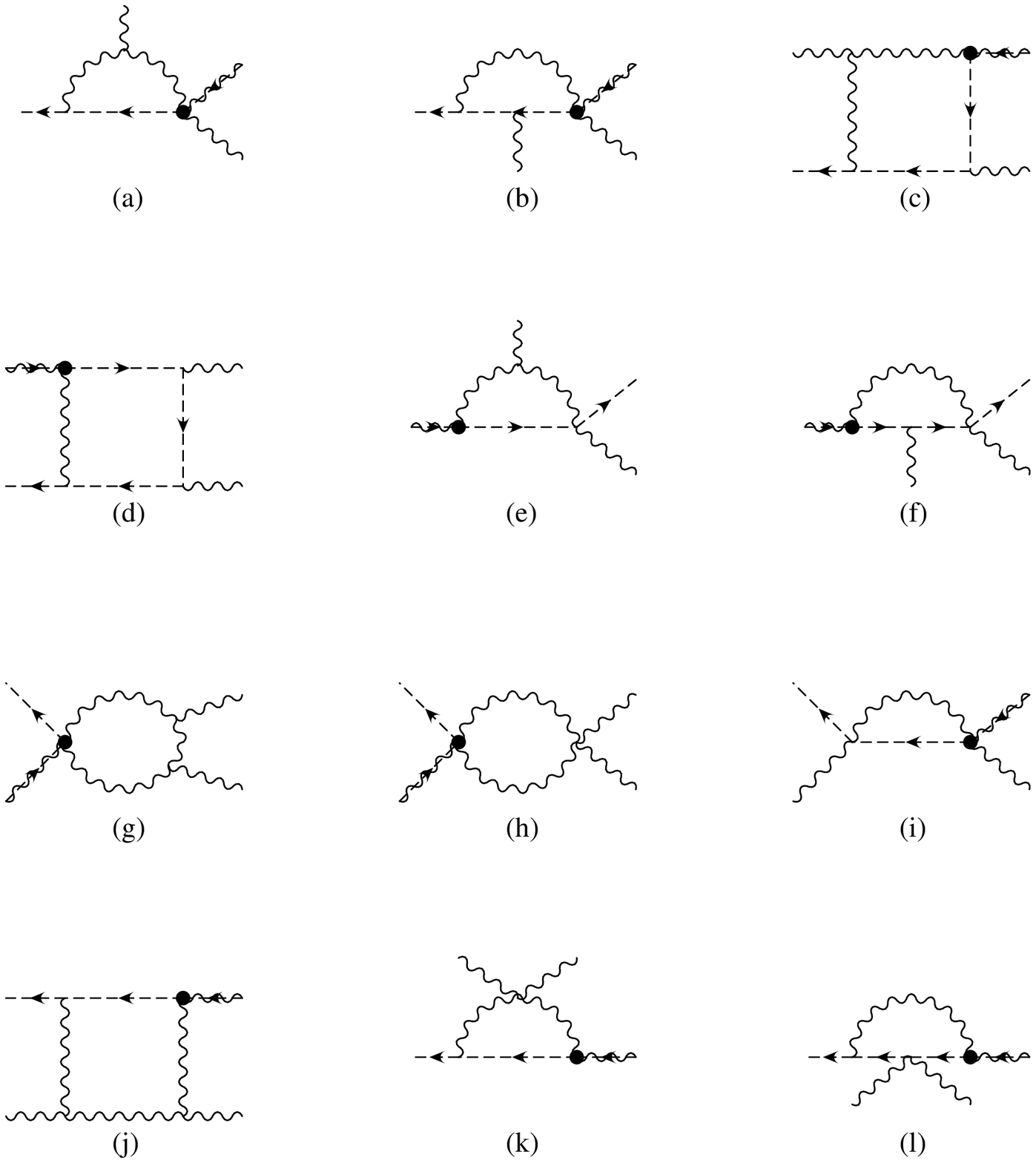}}
\inparg
{\it \noindent Fig. 7: Diagrams with two gauge, one scalar and one 
auxiliary line; the dot represents the position of a $C$.}
\medskip
\outparg
\vfill
\eject
\bigskip
\epsfysize= 5in
\centerline{\epsfbox{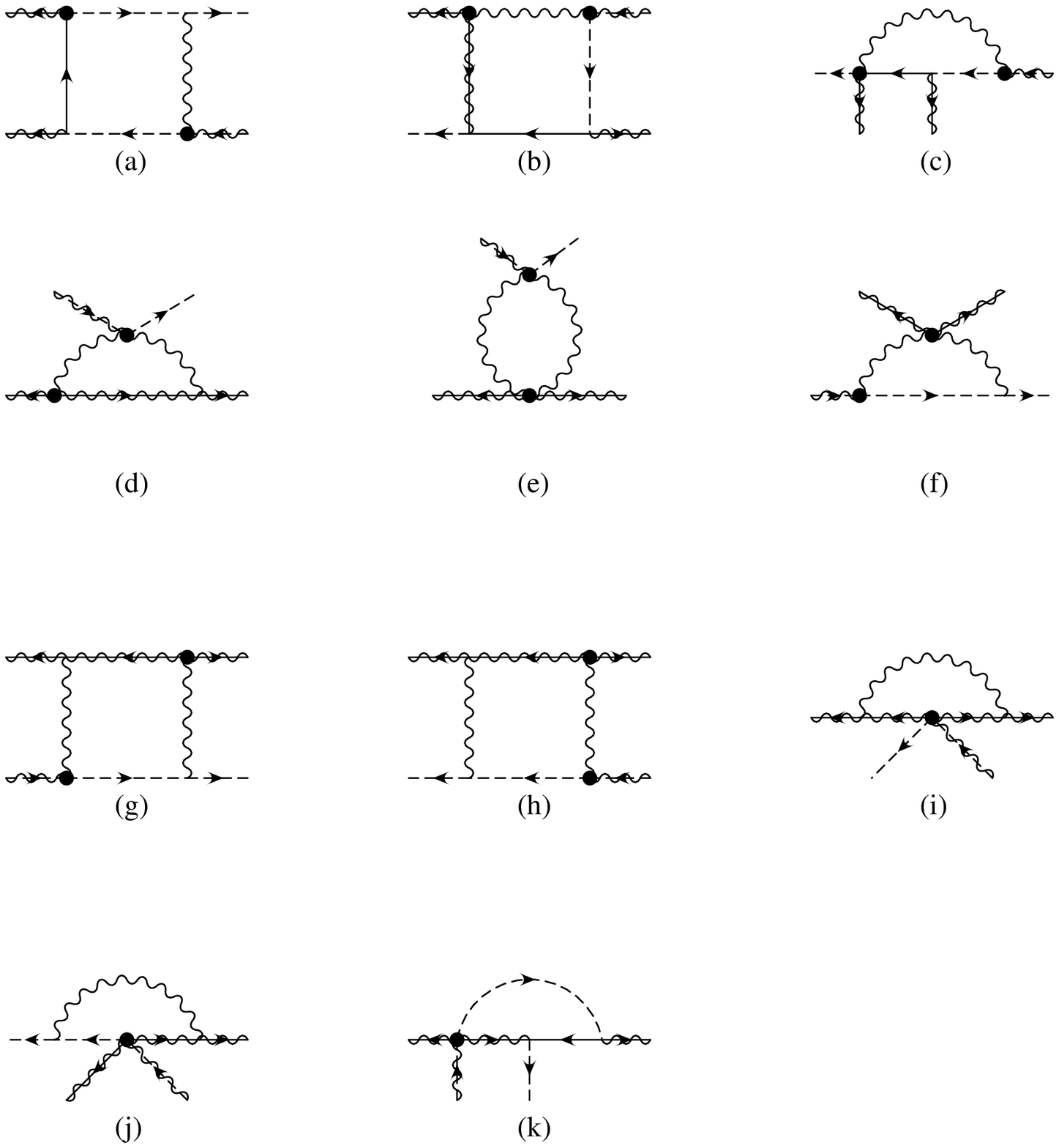}}
\inparg
{\it \noindent Fig. 8: Diagrams with two gaugino, one scalar and one
auxiliary line; the dot represents the position of a $C$ or a $|C|^2$.}
\medskip
\outparg

\centerline{{\bf Acknowledgements}}\nobreak

DRTJ was supported by a PPARC Senior Fellowship, and a CERN Research 
Associateship, and was visiting CERN while most of this work was done.
LAW was supported by PPARC through a Graduate Studentship. One of us
(DRTJ) thanks Luis Alvarez-Gaum\'e for a conversation. 

\listrefs
\bye